\documentclass[trackchanges]{aastex631}
\usepackage{natbib}
\usepackage{comment}
\usepackage{todonotes}
\usepackage{graphicx}

\def\ha{{H$\alpha$}}
\def\hb{{H$\beta$}}
\def\sii{{[\ion{S}{2}]}}
\def\oi{{[\ion{O}{1}]}}
\def\hii{{H~II}}

\def\VEL{\:{\rm km\:s^{-1}}}

\def\OIGS{\:{\rm ergs\:cm^{-2}\:s^{-1}\:\AA^{-1}}}

\def\OiiiL{[\ion{O}{3}] $\lambda\lambda 4959,5007$}
\def\SiiL{[\ion{S}{2}] $\lambda\lambda 6716, 6731$}
\def\NiiL{[\ion{N}{2}] $\lambda\lambda 6548, 6583$}
\def\OiL{[\ion{O}{1}] $\lambda\lambda 6300, 6364$}
\def\SiiiL{[\ion{S}{3}] $\lambda 9069$}

\def\Oi{[\ion{O}{1}]}
\def\Siii{[\ion{S}{3}]}
\def\Sii{[\ion{S}{2}]}
\def\Nii{[\ion{N}{2}]}
\def\Oiii{[\ion{O}{3}]}

\def\H~II{\ion{H}{2}}

\newcommand{\MSOL}{\mbox{$\:M_{\sun}$}}

\newcommand{\EXPU}[3]{\mbox{\rm $#1 \times 10^{#2} \rm\:#3$}}  
\newcommand{\POW}[2]{\mbox{$\rm10^{#1}\rm\:#2$}}

\shorttitle{SNRs in M83}
\shortauthors{Long et al.}
\graphicspath{{./}{figs/}}

\begin{document}

\title{Supernova Remnants in M83 as Observed with MUSE}

\correspondingauthor{Knox S. Long}
\email{long@stsci.edu}

\author[0000-0002-4134-864X]{Knox S. Long}
\affil{Space Telescope Science Institute,
3700 San Martin Drive,
Baltimore MD 21218, USA; long@stsci.edu}
\affil{Eureka Scientific, Inc.
2452 Delmer Street, Suite 100,
Oakland, CA 94602-3017}

\author[0000-0003-2379-6518]{William P. Blair}
\affil{The William H. Miller III Department of Physics and Astronomy, 
Johns Hopkins University, 3400 N. Charles Street, Baltimore, MD, 21218; 
wblair@jhu.edu}

\author[0000-0001-6311-277X]{P. Frank Winkler}
\affil{Department of Physics, Middlebury College, Middlebury, VT, 05753; 
winkler@middlebury.edu}

\author{Lorenza Della Bruna}
\affil{The Oskar Klein Centre, Department of Astronomy, Stockholm University, AlbaNova, SE-10691 Stockholm, Sweden}

\author{Angela Adamo}
\affil{The Oskar Klein Centre, Department of Astronomy, Stockholm University, AlbaNova, SE-10691 Stockholm, Sweden}

\author{Anna F. McLeod}
\affil{Centre for Extragalactic Astronomy, Department of Physics, Durham University, South Road, Durham DH1 3LE, UK}
\affil{Institute for Computational Cosmology, Department of Physics, University of Durham, South Road, Durham DH1 3LE, UK}

\author{Phillippe Amram}
\affil{Aix Marseille Univ, CNRS, CNES, LAM, Laboratoire dAstrophysique de Marseille, Marseille, France}



\begin{abstract}

{Here we describe a new study of the SNRs and SNR candidates in nearby face-on spiral galaxy M83, based primarily on MUSE integral field spectroscopy.   Our revised catalog of SNR candidates in M83 has 366 objects, 81 of which are reported here for the first time.  Of these, 229 lie within  the MUSE observation region, 160  of which  have spectra with [S~II]:\ha\ ratios exceeding 0.4, the value generally accepted as  confirmation that an emission nebula is shock-heated. Combined with 51 SNR candidates outside the MUSE region with high  [S~II]:\ha\ ratios, there are  211 spectroscopically-confirmed SNRs in M83,  the largest number of confirmed SNRs in any external galaxy.  MUSE's combination of relatively high spectral resolution and broad wavelength coverage has allowed us to explore two other properties of SNRs that could serve as the basis of future SNR searches.  Specifically, most  of the objects identified as SNRs on the basis of  [S~II]:\ha\ ratios  exhibit more velocity broadening and lower ratios of [S~III]:[S~II] emission  than H~II regions.  A search for nebulae with the very broad emission lines expected from young, rapidly expanding remnants revealed none, except for the previously identified  B12-174a.   The SNRs identified in M83 are, with few exceptions, middle-aged ISM-dominated ones. Smaller diameter candidates show a larger range of velocity broadening and a larger range of gas densities than the larger diameter objects, as expected if the SNRs expanding into denser gas brighten and then fade from view at smaller diameters than those expanding into a more tenuous ISM.} 
\end{abstract}

\keywords{galaxies: individual (M83) -- galaxies: ISM  -- supernova remnants}

\section{Introduction} \label{sec:intro}

A large fraction of stars more massive than 8\MSOL\ end their lives as supernovae that disrupt all or part of the star, ejecting significant amounts of material at high velocity into the surrounding circumstellar and interstellar medium (CSM/ISM). The interaction of these ejecta in the form of shocks results in emission across a broad range of wavelengths, with the fast primary shock that interacts with the more tenuous medium usually producing X-rays, while secondary shocks driven into the denser phases of the medium emit primarily at ultraviolet, optical, and infrared wavelengths.  Most Galactic supernova remnants (SNRs)  were initially discovered at radio wavelengths, but most extragalactic SNRs have, for reasons of sensitivity and spatial resolution, been identified using narrow-band (interference-filter) imagery  \cite[see, e.g.,][for a review]{long17}.  More specifically, the vast majority of SNRs in nearby galaxies have been identified as nebulae in which the ratio of \SiiL:\ha\ emission is greater than 0.4 to separate them from (bright) H~II regions where the ratio is typically 0.1 -- 0.2.  The physical basis for this distinction is that there is an extended region behind radiative shocks where $\rm S^{+}$ is the dominant species of {sulfur}, while in H~II regions, which are photoionized,  most {sulfur} is found in higher ionization states, primarily $\rm S^{++}$. 

M83 is a well-studied nearly face-on grand-design spiral galaxy with a starburst nucleus and active star formation along its prominent spiral arms.  Because of its  proximity \cite[4.61 Mpc,][]{saha06}\footnote{{We note, in passing, that \cite{bruna21,bruna22} have used a somewhat larger distance of 4.9 Mpc due to \cite{jacobs09}.  We have chosen to use 4.61 Mpc to retain consistency with earlier studies on SNRs in M83.}}and high star formation rate \cite[3-4 $\rm \MSOL\ yr^{-1}$,][]{boissier05},  we and our collaborators have conducted a number of searches for SNRs in {M83. Beginning with \cite{blair04} where 71 SNRs and SNR candidates were identified in M83 using the Dupont telescope at Las Campanas Observatory, members of our team have carried out a series of surveys to compile a more complete and accurate listing of SNRs and SNR candidates in M83, primarily based on optical criteria \citep{dopita10,blair12,blair14}. Currently, the most complete published listing was provided by  \cite{williams19} and includes 278 SNRs and SNR candidates.\footnote{\cite{williams19} list a total of 307 objects, but we have removed 29 objects from that list; see below.}  All of these had appeared in earlier publications.  }

Many of the SNRs appear to have X-ray counterparts \citep{long14} and a significant number are associated with radio sources \citep{russell20}. We have previously confirmed 103 of the brighter [S~II] imaging-selected candidates (out of 118 observed) to have spectroscopic \SiiL:\ha\  ratios greater than 0.4 using Gemini/GMOS \citep{winkler17}. Although the remaining objects observed spectroscopically did not satisfy the standard \sii:\ha\  criterion, \cite{winkler17} suggested that nearly all were still likely SNRs, based on other criteria, such as having coincident X-ray sources and/or significant \oi\ emission.  From an analysis of the stellar populations in the vicinity of the SNRs \citep{williams19}, most are likely the result of core-collapse explosions, as one might expect given the relatively high star formation rate in M83 \citep{boissier05}.  The vast majority of the SNRs that have been observed spectroscopically appear to be evolved ISM-dominated SNRs, based on characteristics of the emission lines observed.   Only two SNRs, one associated with SN1957D \citep{long89,long12} and one object, B12-174a,  that appears to be the remnant of a missed  SN from the last century \citep{blair15}, have lines with the ($> 1000 \VEL$) velocity widths that are seen in very young Galactic SNRs.

Here we describe an analysis of a new homogeneous set of data on SNRs and SNR candidates in the M83, using the Multi Unit Spectroscopic Explorer (MUSE) on the VLT \citep{bacon10}. These observations provide a uniform set of spectro-imaging data covering a significant fraction of the inner bright disk of M83,  
allowing us to sample all of the SNR candidates in the observed region (both those with and without previous spectra), while also providing more extended spectral coverage than was previously available. Combining the MUSE results with previous data, M83  has the most spectroscopically confirmed SNRs of any galaxy studied to date.

The remainder of this paper is organized as follows: In Sec.\ 2, we describe the data used.  In Sec.\ 3, we compile a new catalog of SNRs and SNR candidates, including a revisit to earlier candidate lists as well as adding new MUSE and other previously unpublished candidates. In Sec.\ 4, we discuss the methodology used to extract the MUSE spectra and compare to previous spectroscopy that overlaps the MUSE mosaic region.  In Sec.\ 5, we present an overview of the MUSE results, and in Sec.\ 6, we discuss the limitations that have developed in applying the primary [S~II]:\ha\ ratio criterion for separating SNRs from \hii\ regions.  In Sec.\ 7, we highlight several individual objects of special interest, and in Sec.\  8,  we investigate global trends in the MUSE spectra ,and in Sec.\ 9, we compare the spectra to  shock model grids.  We summarize our findings in  Sec.\ 10.

\section{Data and Data Reduction}

\begin{figure}
\plotone{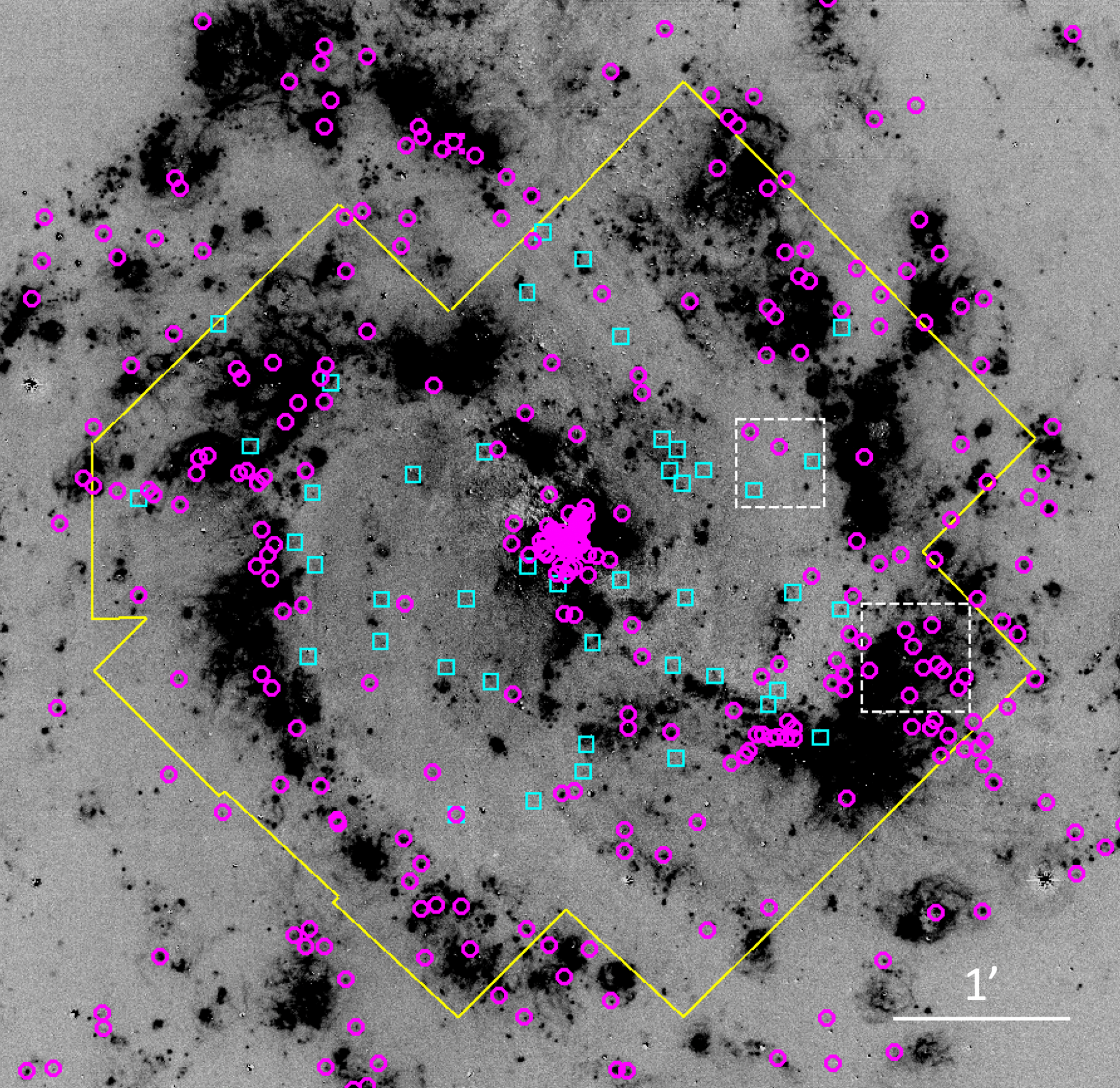}
\figcaption{Magellan continuum-subtracted \ha\ image of M83 showing the region covered by the MUSE mosaic in yellow.  Magenta circular regions show previous SNR candidates, and cyan square regions show the locations  of new  SNR candidates identified with MUSE.  The MUSE candidates are primarily but not exclusively low surface brightness nebulae discovered in the sampled interarm regions.  White dashed boxes show the locations of the regions enlarged in Fig.\ 2 (upper box) and Fig.\ 7 (lower box). A 1\arcmin\ scale bar is shown at lower right.
\label{fig:overview}}
\end{figure}

MUSE is an integral-field spectrograph mounted on ESO's VLT in Chile. For the purposes of this paper, we use data obtained from various ESO programs:  096.B-0057(A), 0101.B-0727(A) (PI Adamo), 097.B-0899(B) (PI Ibar) and 097.B-0640(A)(PI Gadotti).   These data covered the spectral range 4600 \AA\  to 9380 \AA\  with a spectral resolution of about 2.3 \AA, which corresponds to a velocity resolution of about 115 $\VEL$ at \ha. Individual exposures cover a field of 1\arcmin x 1\arcmin; each spaxel is 0.2\arcsec\ $\times$ 0.2\arcsec\ in extent.  As described in detail by \cite{bruna21}, these data have been reprocessed into a single large mosaicked data cube with accurate astrometry and calibration covering the region shown in Fig.\ \ref{fig:overview}.
The median size of point sources in the data cube is 0.7\arcsec\ (FWHM) at 7000 \AA, although there are some slight variations as a function of position and wavelength.  To simplify our analysis in the wavelength range 4600 to 7760 \AA, we have used versions of the data where a stellar continuum has been subtracted using the pPXF fitting code \citep{cappellari04, cappellari17} and the EMILES simple stellar populations \citep{vazdekis16} as described in \citet{bruna21}.  This produced pure emission-line versions of the data cube that made analysis of the nebular emission easier, especially for \hb,  where the stellar continuum often has fairly obvious stellar absorption features.  

M83 has been well observed previously, and these earlier data sets complement the MUSE data is several ways.  Magellan imaging (see Fig.\ \ref{fig:overview}) in excellent (0\farcs 4-0\farcs 5)  seeing \citep{blair12} and HST WFC3 multiband imaging \citep{dopita10,blair14} cover the entire area observed with the MUSE mosaic.  \cite{soria20} describe the detailed HST and multiwavelength characteristics of the complex nuclear region. These higher spatial resolution images can help us understand when the MUSE data may be limited by crowding from nearby sources.  Likewise, Gemini GMOS spectroscopy of objects in common with MUSE  \citep{winkler17} provides confidence in these new results, which extend to objects much fainter than observed with GMOS.

\section{An updated catalog of SNRs and SNR candidates in M83}

{As we have continued to inspect the HST and Magellan data in the context of other projects, such as our recent radio study of M83 \citep{russell20}, we have identified a few additional objects of interest that have not yet appeared in publication.  As part of the current project, we also identify a number of new SNR candidates with MUSE (see next subsection).  Hence, a goal of this paper is to produce a new baseline catalog of SNRs and SNR candidates for use going forward that includes all of these objects.}

In the following subsections, we first discuss the set of new SNR candidates identified uniquely with MUSE from within the MUSE footprint shown in Fig.\ \ref{fig:overview}.   
Then we discuss a few special cases as well as {the previously unpublished candidates, many from the nuclear region which had been set aside in earlier studies.}   In creating the revised list, we {remove from consideration the }
\Oiii-selected list of candidates from \cite{dopita10} (5 objects) or \cite{blair12} (their Table 3) unless existing data has already confirmed the objects as good SNR candidates; most of the \Oiii-selected objects have been shown to be Wolf-Rayet nebulae or other compact H~II regions and are no longer considered viable SNR candidates. Some 24 such objects were listed in the \cite{williams19} table, as were 5 historical SNe that do not have known SNR counterparts. These 29 objects have been removed from the new catalog.

\subsection{A MUSE Search for Additional Supernova Remnants \label{sec:more}} 

\begin{figure}
\plotone{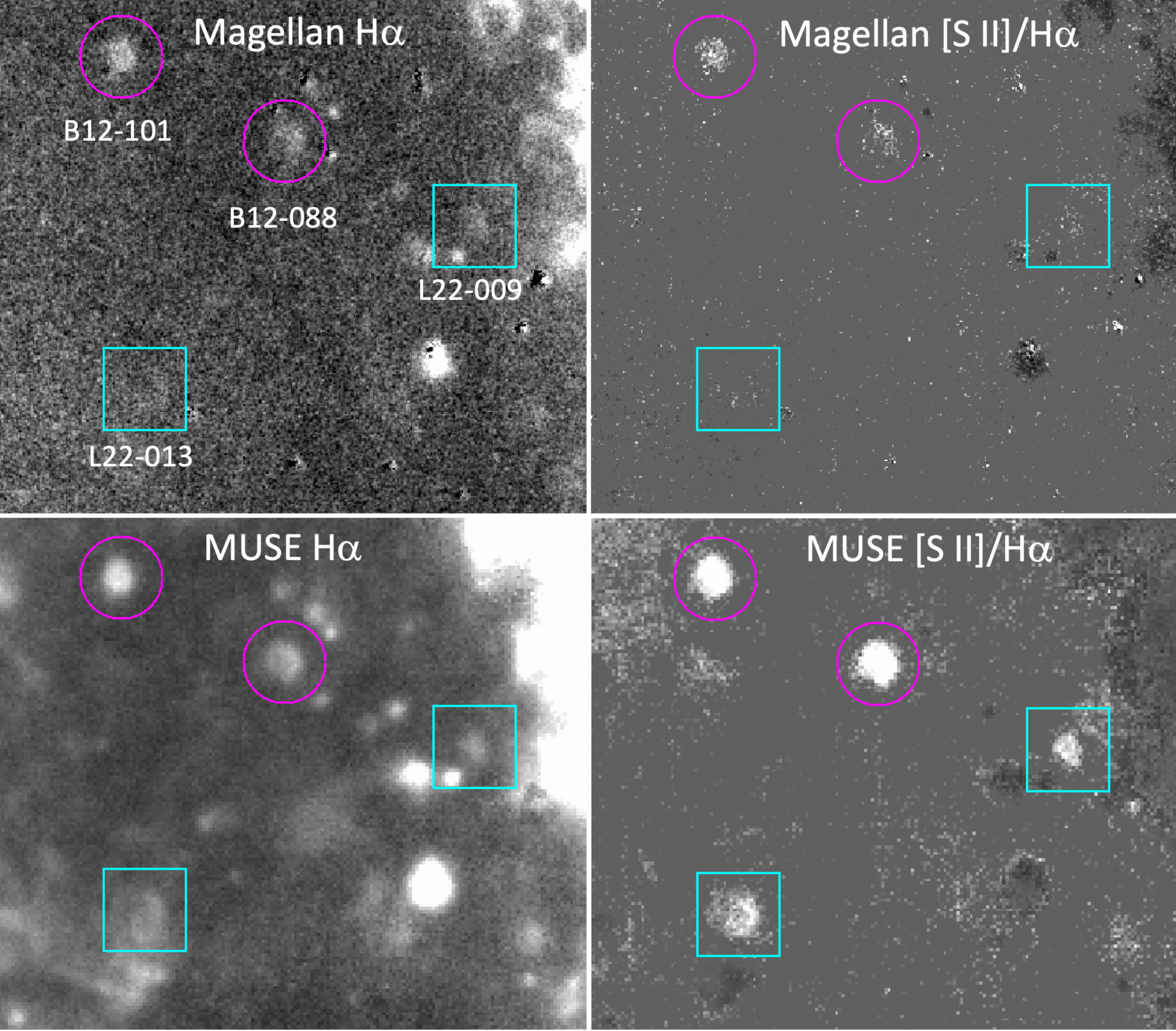}
\caption{A comparison of \ha\ images and [S~II]:\ha\ ratio maps for a 30\arcsec\ region in M83.  The top two panels show previous Magellan IMACS data \citep{blair12} and the bottom two panels show the corresponding data from MUSE. The scaling on the ratio maps is from 0 (black) to 2.0 (white).  Thus, lighter values indicate elevated values of the ratio.  The magenta circles mark two previously identified SNRs (B12-101 and B12-088) while the cyan squares indicate two additional SNRs identified with MUSE (L22-013 and L22-009).  The improved exposure depth and higher S/N in the MUSE data are evident for this relatively sparse inter-arm field. The previous objects have elevated ratios in the Magellan data, but even higher ratios in the MUSE data.  The new objects stand out in the MUSE ratio and have identifiable emission counterparts, but are too marginally detected in the Magellan data to have been identified previously as SNR candidates.  For scale, the square regions are 5\arcsec\ on a side. \label{fig:muse_new}}
\end{figure}

As noted earlier, most SNRs in nearby galaxies have been identified using interference filter imagery.  Usually the ``\ha'' filter passes a significant amount of one or both of the adjacent \NiiL\ lines.  Typically, one forms a ratio image by first subtracting a scaled continuum image from the \ha\  and \sii\ images, and then dividing to create the ratio image that one searches for emission nebulae with higher \sii:\ha\ ratios than \hii\ regions in the image.  Much cleaner ratio images can be created with a  high (spectral) resolution integral spectrograph like MUSE, since one can exclude the effects of \Nii\ contamination of \ha\ and one can subtract a continuum on either side of the emission line. In our case, we chose to create the ratio images from using 10 \AA\  wavelength bands for the two {sulfur} lines and \ha, subtracting adjacent continuum bands, and then creating the ratio image.  By way of example, a portion of the ratio image is displayed in in Fig.\, \ref{fig:muse_new}. 

Many of the previously known SNRs were of course prominent in this ratio map, as shown by the magenta regions in Fig.\, \ref{fig:muse_new}, where the two previously identified objects are detected at significantly higher \sii:\ha\ ratios than in the Magellan data. However, there  were a number of additional locations in the MUSE ratio map that appeared to have elevated ratios; the cyan regions in Fig.\, \ref{fig:muse_new} are two such objects.  Indeed, a visual search over the entire MUSE mosaic  revealed 44 such objects of interest that passed the following vetting procedure:  First, we confirmed that each region showing an elevated ratio had an  identifiable \ha\ counterpart by displaying the MUSE \ha\ image alongside the ratio map in ds9.  We then verified that there was, as shown in Fig.\, \ref{fig:muse_new}, evidence of the nebulae in our high spatial resolution ($\sim 0\farcs 5$ FWHM) Magellan IMACS emission-line image, but at a level too low to confirm their identity with the previous data.  A major reason for this vetting was to avoid the possibility of selecting ill-formed regions of diffuse ionized gas (DIG) to be SNR candidates.  These checks combined make us reasonably confident in the selection of these new objects as SNR candidates, i.e., isolated nebulae with higher than normal [S~II]:\ha\ ratios, and not DIG.  They are included in our new catalog.  The object B14-48 that \cite{winkler17} removed from consideration was already removed from the \cite{williams19} list.

In Sec.\  \ref{sec:extraction}, we discuss the extraction of MUSE spectra at the positions of the SNRs and SNR candidates in M83, and  confirm that many, but not all, of these new MUSE  candidates have [S~II]:\ha\ ratios above 0.4, as anticipated from the imaging.  On average, objects in this group have derived surface brightnesses toward the lower end of the SNR sample.  One could imagine pushing observations even lower in surface brightness to possibly find additional SNR candidates, but at some point, it becomes impossible to define an actual emission nebula from a slightly brighter patch of DIG in the general ISM.  Since the spectral characteristics of the DIG can mimic that of slow shocks,
e.g., weak [O~III] and an enhanced \sii:\ha\ ratio 
{\citep{galarza99,poetrodjojo19,bruna21},} it becomes increasingly difficult to distinguish true lower surface brightness SNRs from DIG even with the sensitivity provided with MUSE/VLT.  This is discussed further in Sec.\  \ref{sec:faint}.

\subsection{Other New SNR candidates and Special Cases}

Our ongoing inspection of the Magellan and HST data sets continues to uncover a handful of objects of potential interest that have not been reported previously. Eleven such objects found in the original Magellan imaging data \citep{blair12} are now included in the catalog. 
Spectroscopy (combined with multiwavelength considerations) occasionally has allowed some earlier objects to be eliminated from further consideration; \cite{winkler17} for example, concluded the object B14-48 had a low ratio and no other interesting characteristics that pointed to shock heating, so it was dropped.

The largest number of previously unpublished candidates arises from the HST imaging in the nuclear region.  \cite{blair14} describe the use of the near-IR line of [Fe~II] $\lambda$1.644 $\mu$m as a new shock diagnostic and published a number of M83 SNRs showing this line. However, the situation in the complicated nucleus was set aside and not reported in that analysis.  For completeness here, we include 20 new  SNR candidates selected using the presence of strong [Fe~II] emission but little or no optical emission, the vast majority of which are in the nuclear region. Most of these objects are seen in projection against dust lanes and are thought to represent heavily reddened SNRs seen primarily in [Fe~II].\footnote{There are nine other previously-published SNRs in the catalog selected on the basis of strong [Fe~II] emission, so 29 such objects total.  26 of the 29 [Fe~II] objects are within the MUSE mosaic.}  

Among the SNR candidates, detailed studies of individual objects have identified a few as microquasars (or candidates) whose spectra also involve shocks.  The microquasar MQ1=D10-N-16 \citep{soria14} is found just NE of the bright nuclear region, and more recently \cite{soria20} suggested that two closely spaced SNRs, B12-096 and B12-098, may be two shock-heated lobes of another single microquasar.  This latter paper highlighted a handful of other M83 SNR candidates that had peculiar morphology, but none of them panned out as microquasar candidates.  We have maintained these three objects in the catalog because they display characteristics of shock heating, but we annotate them for future reference.

\subsection{The New Catalog \label{sec:catalog}}

Our current list of SNRs and SNR candidates in M83 now consists of {366} objects as indicated in Table \ref{snr_master}.   The catalog includes the 278 nebulae listed by \cite{williams19}. To this, we have added a total of {87} other {nebulae, 81 of which are reported here for the first time: 44 from our search of the MUSE data, and 37 new previously unpublished candidates from re-inspection of the earlier imaging data of the M83 nucleus.  Six other objects previously published by \cite{dopita10} or \cite{blair14} that were considered too marginal to be included in the \cite{williams19} list have been reinstated here for completeness.  This new catalog} comprises the largest optical SNR sample available for any single galaxy.


For each object, we list a recommended name, coordinates, an estimated diameter, the distance of the object from the nucleus, the identification of any X-ray sources that are co-spatial with the SNR candidate, an indication of whether the object was included in the list of \cite{williams19}, seen first in the MUSE data, or has been based on further inspection of the HST or Magellan data and whether GMOS or MUSE spectra exist.  For some objects, a comment is added.  The diameters were measured using the HST images wherever possible; for a relatively small number of low surface brightness, large diameter objects, the Magellan or MUSE images were used.  The galactocentric distances assume an inclination and position angle of  24\degr\ and 45\degr, respectively \citep{talbot79}. The X-ray source list utilized is that of \cite{long14}.  Where possible we have used the same names as cited by  \cite{williams19}; completely new objects, L22-001, etc are in right ascension order (as is the entire table).  

\section{Extraction of the MUSE spectra \label{sec:extraction} }

Of the {366} objects in Table \ref{snr_master}, 229 lie within the area outlined by the MUSE mosaic.   The primary difference between the sample of objects in and out of the MUSE region  is that the MUSE sample contains all of the SNR candidates in the nuclear starburst of the galaxy and all of the objects that were identified on the basis of strong [Fe~II] emission.  The diameter distribution of the  MUSE sample skews somewhat toward smaller diameter objects; the median (average)  diameter of the objects in and out of the MUSE region is {20  (22) pc and 25 (27)} pc, respectively, a difference we consider to be inconsequential.

\begin{figure}
\plotone{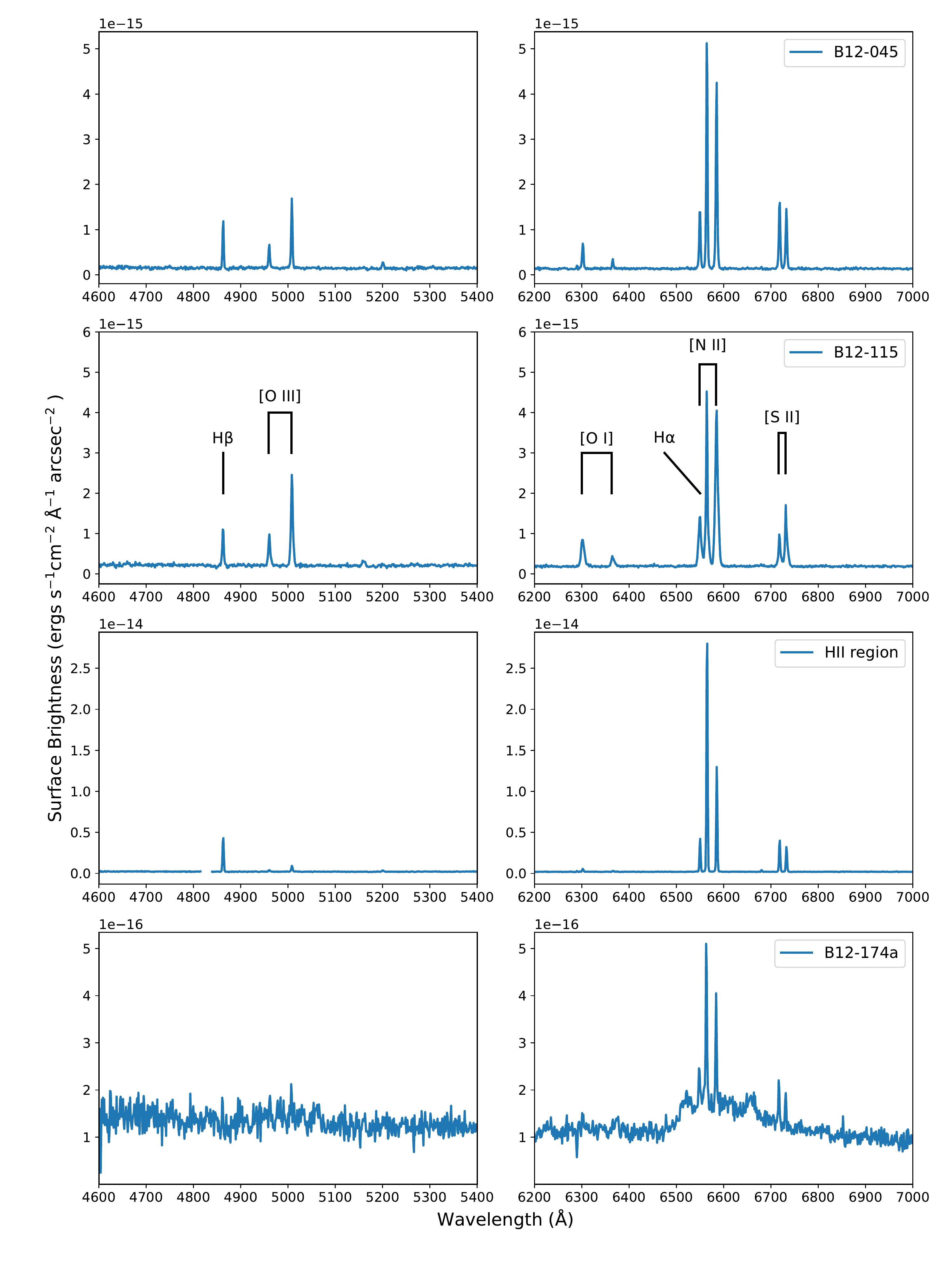}
\figcaption{Some example background-subtracted MUSE spectra. {The plots have been adjusted to show the lines at rest wavelength.}The top two panels show typical SNR spectra, including \OiL\ in addition to the other red shock indicators (note the high density indicated by the [S~II] ratio for B12-115). The third panel shows a typical but comparably faint \hii\ region, for which the forbidden lines are much weaker and \Oiii\ is very weak. Finally, the bottom panel shows the very young SNR B12-174a \citep{blair15}, which shows a very broad emission hump in the red. (See text Sec. 7 for details.)  Using MUSE off-band images adjacent to \ha\ + [N~II], we were unable to find any other objects with this sort of broad emission profile. 
\label{fig:example}}
\end{figure}

As is apparent from an inspection of Fig.\ \ref{fig:overview}, many of the SNRs and candidates are located in or near other nebulosity.  Consequently, one must allow for the effects of possible contamination when extracting spectra.  With such a large number of objects, setting individual regions for background subtraction is impractical.  Hence, we have experimented with a number of more automated approaches for selecting and subtracting background from the spectra. In the analysis that follows, we have used the following procedure.  We first extracted the average spectrum from all the spaxels contained within the region centered on the object, with an angular radius of the extraction region given by

\begin{equation}
\theta= \sqrt{\theta_{snr}^2 + \theta_{psf}^2}
\end{equation}

\noindent
where $\theta_{snr}$ is the angular radius of the SNR measured from high resolution HST images of the object \cite[cf.][]{blair14} and $\theta_{psf}$ is 0.35\arcsec.\footnote{As described by \cite{bruna21}, there is some variation in the size of the point spread function over the field.  We did not take these changes into account, but we did confirm, by testing different values of the point spread function, that the exact value of the point spread function made very little difference in the final results.}  To obtain the local background for subtraction, we first located the spaxel with the lowest \ha\ flux within 5\arcsec\ of the SNR.  We then extracted an average background spectrum from the spaxels within 1.5\arcsec\ of that lowest flux spaxel.  The net spectrum (with units $ \OIGS \rm spaxel^{-1}$) was then obtained by  straight subtraction after scaling to the same areal size as the object.
As an indication of the resulting data quality, a selection of spectra is shown in Fig.\  \ref{fig:example}.

Applying this procedure allowed us to treat the spectral extractions in an automated way, and in general this procedure worked well.  (See below for comparison to objects that had prior spectra with Gemini GMOS.)  However, we do note that, particularly for faint objects, a different choice of background region would result in different line ratios for individual objects.  However, this criticism would apply to almost any choice of background, even one made ``by eye''.

In addition to the spectra of the SNR and SNR candidates, we extracted spectra for 188 other emission nebulae, listed in Table \ref{h2_master},  that were more likely to be H~II regions. These objects were selected by displaying the MUSE \ha\ image and randomly selecting small isolated patches of \ha\ emission that spanned the surface brightness regime of the SNR candidates in the \cite{williams19} list. In choosing the comparison \hii\ regions in this way, we have avoided the bias that might occur from selecting bright \hii\ regions that are less typical.  We also took care to select objects scattered throughout the portion of M83 covered by the MUSE data cube so as not to be biased by local effects. Background subtraction for these objects was performed in a manner similar to that described above for the SNRs.  

With the spectra in hand, we extracted central wavelengths, fluxes and FWHM for the lines in the various spectra using a custom built {\sc Python} script, assuming that the lines had Gaussian shapes.  Our purpose-built fitting routine makes use of the {\textsc{curve\_fit}} module of {\textsc{scipy.optimize}}, which returns not only the best fit values but also the covariance matrix.  We have used the covariance matrix to establish 1$\sigma$ error estimates under the assumption that the errors for the various values were uncorrelated.
For the doublets, \OiiiL, \OiL, and \SiiL, we fixed the separation of the lines and fit a single FWHM to both lines, which particularly in the case of \Oi\ and \Oiii\ produced more accurate results for the weaker of the lines in the doublet.  \hb\ and [S~III] $\lambda$9069 were treated as singlets.  

After some experimentation, we decided to treat the \NiiL\ and \ha\ complex as a single system, fixing the separation between the lines, and using a single FWHM. In addition to visually inspecting the results, we carried out a number of checks to validate the fits, including in the cases of typically weaker lines (such as \hb, [O~III], [O~I] and  \Siii)  that the fitted wavelengths and broadening of the feature were consistent with those of the stronger lines of [N~II], \ha\ and [S~II].  To establish upper limits to the flux in cases where the line was not clearly evident, we created versions of the spectra where we artificially added a line of known strength and FWHM at the position of a line.   We fit the modified spectrum and used the error in the flux to estimate a 1$\sigma$ upper limit on a line at the position of the emission lines.  
 
The results are presented in Tables \ref{snr_spectra} and \ref{h2_spectra} for the SNR candidates and \hii\ regions, respectively.  As is traditional, we have reported the line strengths relative to a reference \ha\ value, but provide the \ha\ surface brightness in physical units for scaling purposes.  Although we have only presented the FWHM from the \ha\ fits, the FWHM of all the lines were, within the errors, consistent with one another.

\subsection{Comparison to Previous Spectra}

\cite{winkler17} used GMOS on Gemini-South to obtain moderate (3.9\AA\  at \ha) resolution spectra of 118 of the mostly brighter SNR candidates known previously.  Their spectra confirmed that 103 of the objects had \sii:\ha\ ratios of 0.4 or larger.   Of the objects with lower ratios, most (13) showed evidence of emission from \oi\,$\lambda$6300, nine had [Fe\,II]$\lambda$\,1.644 $\mu$m emission, and seven were soft X-ray sources. Many of these objects also resided in complex regions of emission, making background subtraction more inaccurate, which could contribute to low observed ratios.  In the end, \cite{winkler17} argued that all but the aforementioned B14-48 were likely to be SNRs despite their lower observed ratios. 

\begin{figure}
\plottwo{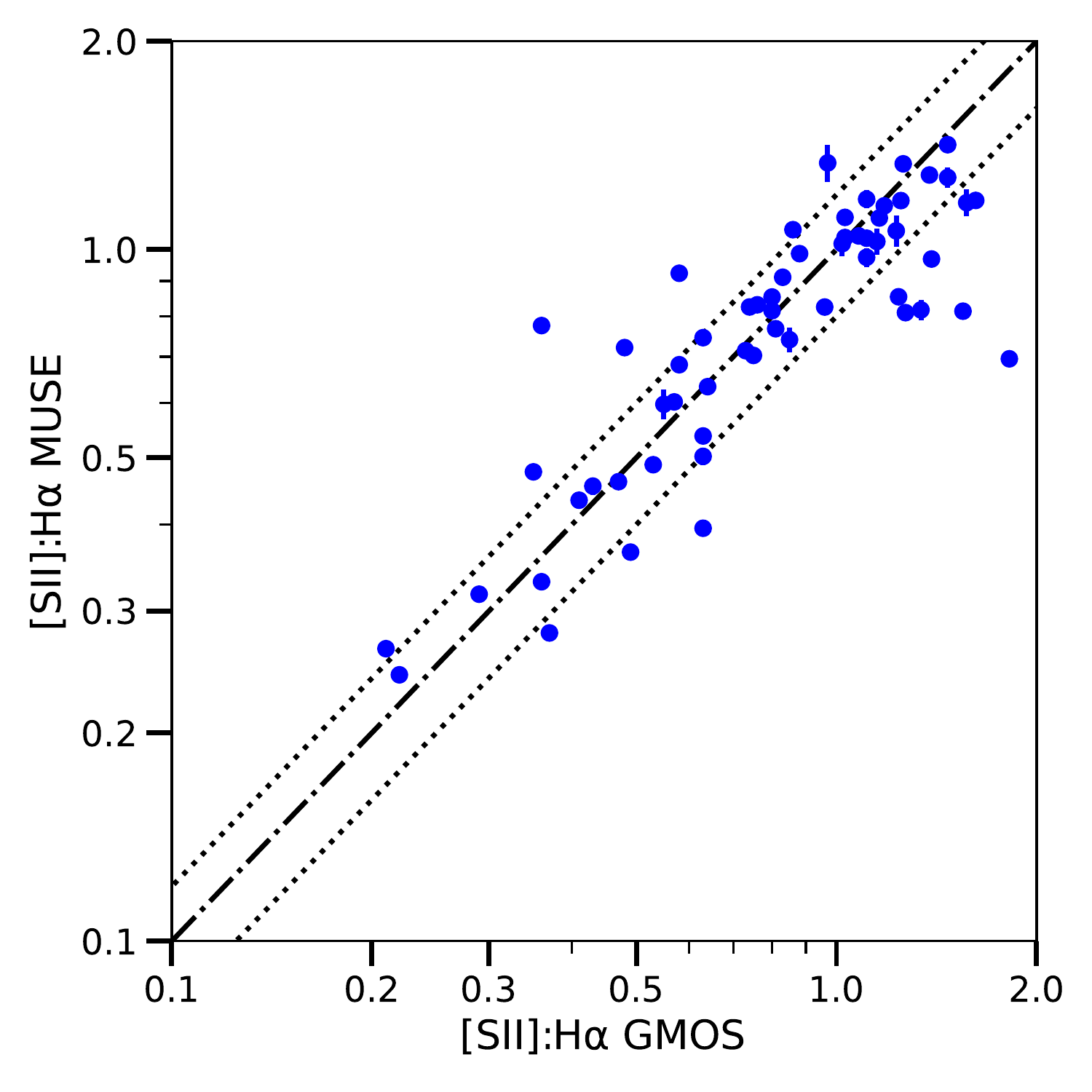}{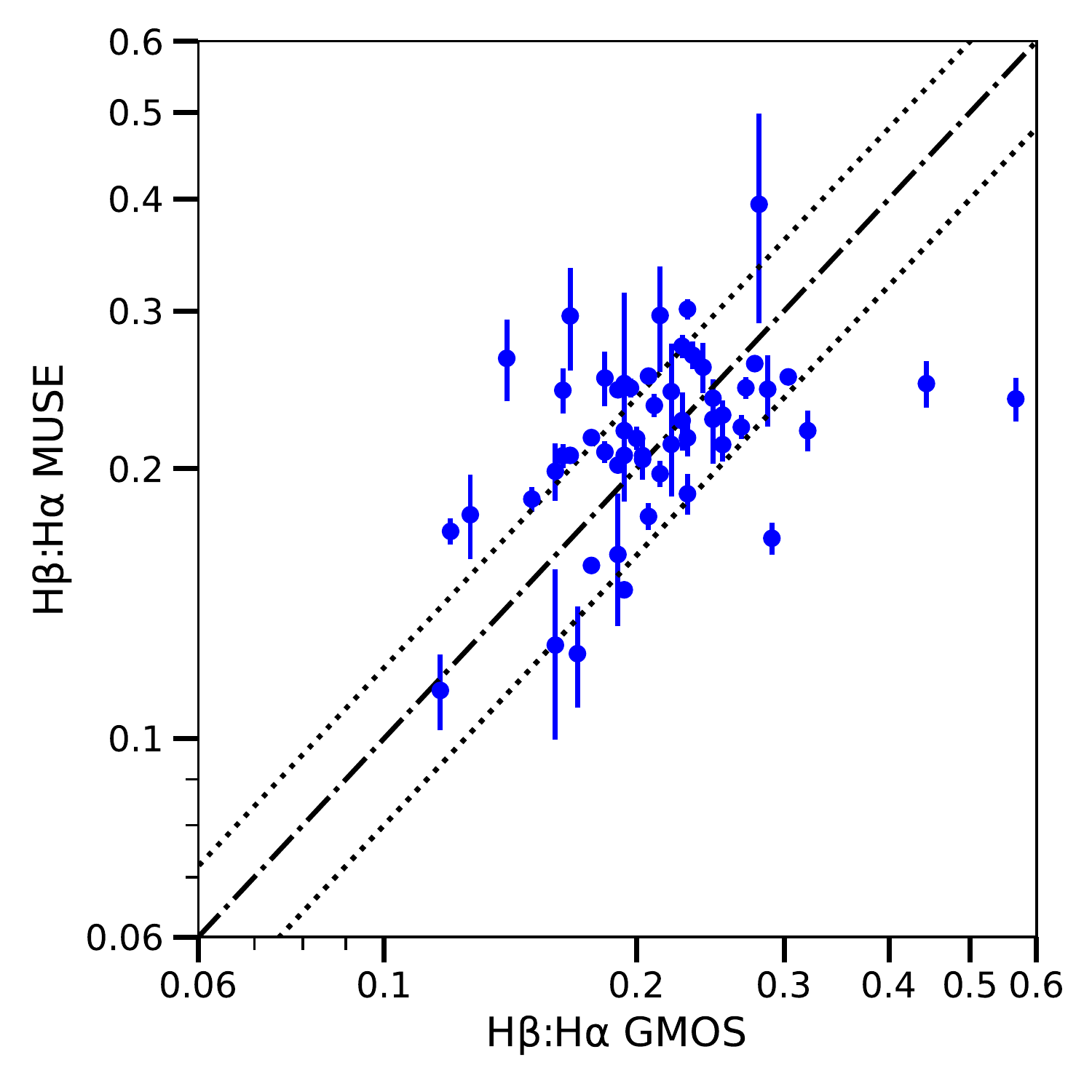}
\caption{Left: A comparison of [S~II]:\ha\ ratios of SNRs observed with GMOS and MUSE.  Right: A comparison of \hb:\ha\ ratios of the SNRs.  Ideally all of values would lie along the 'dot-dash' curves, but modest scatter is present.  The ``dotted" lines show the locus for a 20\% difference between the measurements with GMOS and MUSE and most of the points are encompassed. The main outliers in the left panel show higher ratios in the GMOS data, consistent with better isolation and less contamination in these slitted observations. The weakness of \hb\ (and hence larger uncertainties) accounts for the scatter seen in the right panel. \label{fig:gmos}}
\end{figure}

Of the objects with spectra in \cite{winkler17}, 59 are located within the region studied with MUSE, making a direct comparison possible. As one might expect, and as shown in Fig.\ \ref{fig:gmos},  there is a fairly good correlation between the [S~II]:\ha\  and the \hb:\ha\ ratios measured with the two instruments.  Not surprisingly, the scatter is much larger for  \hb:\ha\ ratios, as \hb\ is typically much fainter than the [S~II] lines.  (Results for [O~III]5007:\hb\ are also similar to that of \hb:\ha.)   Our basic conclusion is that while there are some differences in the individual object line ratios, there is no indication of any systematic issues with the process we have used in extracting and analyzing the spectra.  

Having said this, we note that there does appear to be a tendency for objects that exceed a [S~II]:\ha\ ratio of 0.4 in the MUSE spectra to have even higher values of the ratio in the GMOS spectra.   This would be the sense if the GMOS slit observations were more effectively allowing local contamination to be subtracted. However, the differences could also be due to the fact that in the case of the MUSE spectra, the spaxels used in creating the source and background spectra were created by a mechanical process based on the size of the source and the point spread function, whereas \cite{winkler17}'s  GMOS extractions were carried out using  {source and background regions that were selected individually for each object.}

\section{Results}

{Applying the traditional [S~II]:\ha\ criterion to the objects within the MUSE footprint, we find the following:
Of the 229 objects within the MUSE field that have previously been classified} as SNRs on the basis of narrow-band imagery, 160 have MUSE-determined [S~II]:\ha\ ratios of 0.4 or greater; 187 have ratios greater than 0.3.  By contrast, for the random sample of 188 H~II regions, only seven have a ratio of 0.4 or more, and 26 have a ratio greater than 0.3.  The median value of the ratio is 0.62 for the SNR sample, 0.20 for the H~II region sample.  The latter value is higher than the value of 0.1 that is typically quoted for H~II regions, but that value is more applicable to high surface brightness \hii\ regions.

\begin{figure}
\plottwo{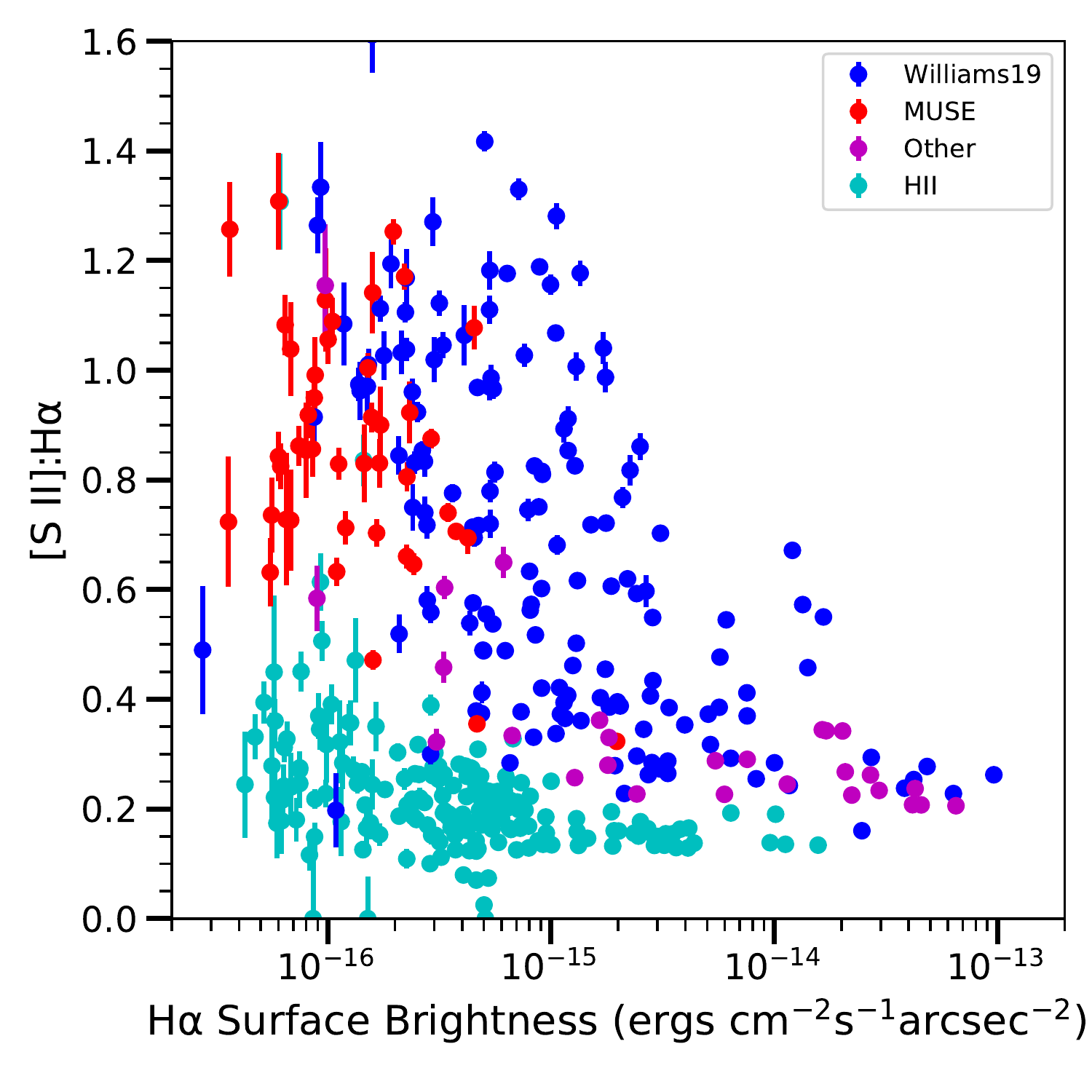}{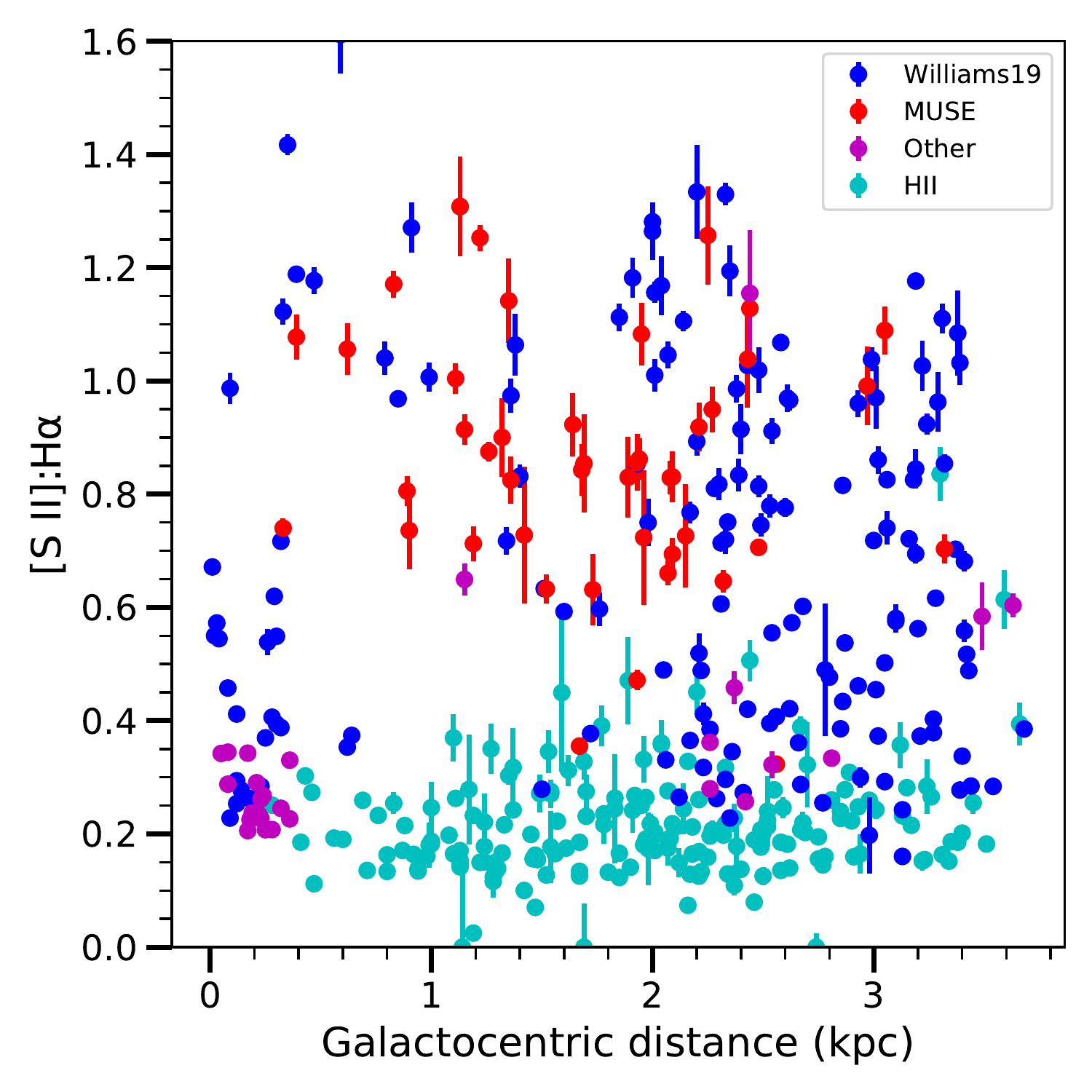}
\figcaption{Left: The MUSE-derived [S~II]:\ha\ ratio as a function of \ha\ surface brightness for objects in our catalogs.  Objects shown in blue were in the \cite{williams19} list of SNRs and SNR candidates while purple shows results for some of the additional candidates described in Sec. 3 of the text.
Also shown in red are the newly identified MUSE SNR candidates which, as expected, trend toward the lower surface brightnesses. Results from the H~II region sample are shown in turquoise.  Right: [S~II]:\ha\ ratio as a function of galactocentric distance.  The 1$\sigma$ errors in the  [S~II]:\ha\ ratios are shown. No particular trends are seen, nor were they expected, in this plot.  Note that the newly identified MUSE SNRs look entirely consistent with the earlier sample.    \label{fig:s2_ratio}}. \end{figure}

Fig.\  \ref{fig:s2_ratio} shows the derived [S~II]:\ha\ ratios as a function of \ha\ surface brightness and as a function of deprojected galactocentric distance for both the SNRs and H~II region samples.  
There is an obvious trend for the SNRs to have higher ratios at lower surface brightnesses, although there is a large dispersion. This is, for the most part, a selection effect that results from how the sample was constructed; it is easier to identify bright objects with elevated ratios than faint ones, so in the case of the faint objects one actually ``needs'' a higher ratio to pick out the object.
At the high-surface-brightness end of the distribution, a number of the SNR candidates have MUSE-derived ratios below 0.4; most of these objects still seem to be elevated relative to H~II regions of comparable surface brightness, but the distinction is less clear.  Most of these objects have soft X-ray counterparts or other indicators that help to solidify the SNR identifications.  Still, this points to a limitation in blindly applying the \sii:\ha\ $\ge$ 0.4 criterion across the board, a topic that is discussed more thoroughly in Sec.\ \ref{sec:limitations} below.

Observationally, it is clear the nebulae selected to be \hii\ regions generally have \sii:\ha\ ratios that are near \sii:\ha\ = 0.2, although there is a trend for lower surface brightness \hii\ regions to have more elevated ratios, even approaching or exceeding the nominal 0.4 ratio discriminant for shock heating.  However, the low surface brightness SNR sample trends toward even higher values of the ratio as well, thus largely maintaining a separation from the \hii\ regions of similar surface brightness.  We thus maintain confidence that most of these faint SNR candidates are viable. 

As shown in the right panel of Fig.\  \ref{fig:s2_ratio}, there are no obvious trends in the [S~II]:\ha\ ratio as a function of galactocentric distance in either the SNR or \hii\ region samples.  There is a tendency for the SNRs in the nuclear region to have systematically lower ratios.  There are 48 candidates with galactocentric distances less than 0.5 kpc; of these only 20 or 42\% have [S~II]:\ha\ ratios greater than 0.4.  There are 181 candidates further from the nucleus; 140 or 77\% satisfy the [S~II]:\ha\ criterion.  This is not unexpected; a number of the SNR candidates in the nuclear region were selected primarily on the basis of the [Fe~II] emission, and were quite faint in \ha.  Additionally, background subtraction is more difficult in the very complex nuclear region.\footnote{As noted earlier, \cite{winkler17}  obtained GMOS spectra showed 103 of 118 SNR candidates with [S~II]:\ha\ ratios greater than 0.4, a success rate of 87\%.  The lower success rate obtained with MUSE is due primarily to the fact that the MUSE sample includes the SNRs in the nuclear region, and secondly to the fact that the MUSE spectra include both bright and faint SNRs in the FOV, whereas objects in the GMOS study were selected at some level to be bright.}

\begin{figure}
\plottwo{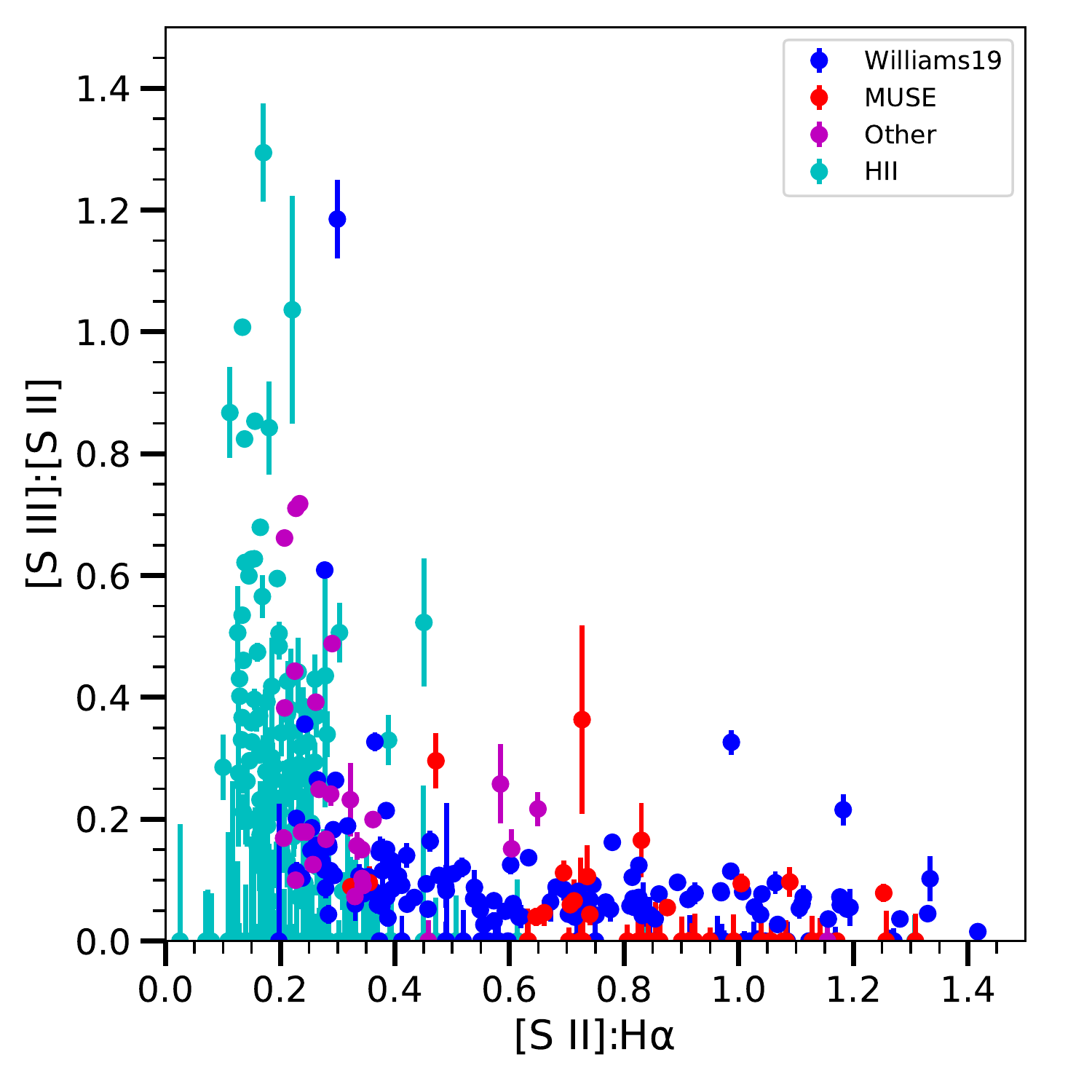}{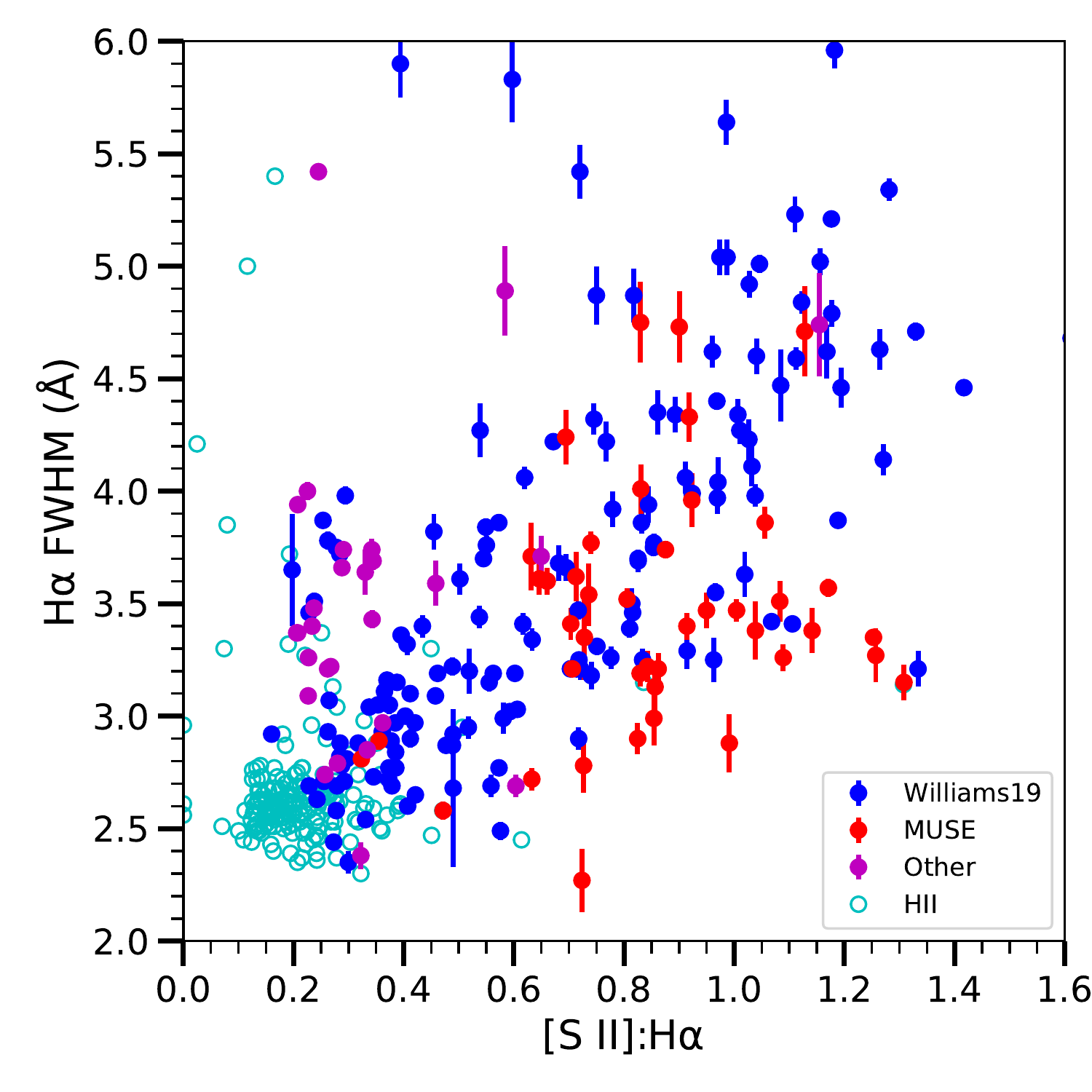}
\figcaption{  Left: [S~III]:[S~II] as a function of the [S~II]:\ha\ ratio. SNRs trend toward low values of this ratio. Objects where no [S III] was detected lie along the x-axis with 1 $\sigma$ upper limits on the ratio.  Right:  The FWHM of \ha\ as a function of the [S~II]:\ha\ ratio. SNRs tend to have broadened line profiles due to bulk motions of the shocked gas. For the SNRs, 1$\sigma$ errors on the fitted FWHM are shown.   Colors are the same as Fig.\ \ref{fig:s2_ratio}.  \label{fig:s2_fwhm}}
\end{figure}

The extended red spectral coverage of the MUSE spectra and the $\sim$100 $\VEL$ spectral resolution allow us to look at these samples in additional ways.  The left panel of Fig.\  \ref{fig:s2_fwhm} uses the \SiiiL\ line in ratio with \sii\ and compares this against the \sii:\ha\ ratio  to show something quite interesting.  As noted earlier, $\rm S^{++}$ is the dominant species of {sulfur} in \hii\ regions, whereas $\rm S^{+}$ is prominent in the recombination and cooling zones of shocks.  As expected, those objects selected as \hii\ regions trend toward significantly higher [S~III]:[S~II] ratios. 
In fact, 58\% of the H~II region sample has a [S~III]:[S~II] ratio greater than 0.1, whereas only 10\% of the SNR sample does.
While [S~III] can be seen in SNR spectra (just like [O~III], depending on shock velocity), the much stronger [S~II] emission from the recombination zone of the shocks forces the SNR [S~III]:[S~II] ratios to be systematically lower.  
The separation is not clean, however and a number of objects are in an overlapping region in the lower left of the plot.  We note that no correction for differential extinction has been made here; a correction would only increase the relative strength of [S~II] and drive sources toward lower [S~III]:[S~II] ratios.  While the near-IR region containing the [S~III] line is not often observed for SNRs, it appears that the [S~III]:[S~II] ratio offers a secondary diagnostic that can help determine the ionization character of uncertain objects. {A possible advantage of the [S~III]:[S~II] ratio as a shock discriminator is that it involves only a single element, and therefore is not sensitive to questions of relative abundance. }

As recently emphasized by \cite{points19}, with sufficiently high spectral resolution, SNRs (at least ones with significant shock velocities) can be separated from H~II regions kinematically on the basis of their line widths.  Indeed, \cite{mcleod21} have used MUSE spectra to confirm a number of SNRs  in NGC300.   The utility of measuring line widths is also evident in the MUSE spectra of M83 SNRs and candidates.
In the right panel of Fig.\  \ref{fig:s2_fwhm}, we show the derived full width at half maximum (FWHM) values for the \ha\ line from our simple Gaussian fits.  
Here, with the exception of a handful of objects, there is good separation between the SNRs, which show signs of broadening, and H~II regions, which effectively scatter around the instrumental resolution of $\sim$2.3 \AA.  For the SNR sample, 68\% of the SNR candidates have \ha\ line widths which exceed 3 \AA, whereas only 5\% of the objects in the H~II region sample do.  The H~II region outliers can be attributed to very low surface brightness objects with poorly determined line widths. The SNRs at the lowest FWHM values either have slow shocks or their spectra are contaminated by H~II emission such that the higher velocity emission is masked. This discriminant works well despite the fact that significant dispersion is present within the SNR sample itself.  We also note that the newly determined MUSE SNR sample (red points) looks similar to the earlier objects from the \cite{williams19} catalog in both of the \Siii:\sii\ and kinematic  diagnostics.

Another well-known shock indicator is the [O~I]:\ha\ ratio, which, as recently emphasized by \cite{kopsacheili20}, should be near zero in photoionized gas but with apparent \oi\ emission in shocked gas.  In M83, as a result of its recession velocity, the [O~I] doublet in M83 is fairly well separated from [O~I] in the airglow.  Two-thirds of the SNR candidates have [O~I]:\ha\ $\ge$  0.1.  Interestingly, there are no SNR candidates with [S~II]:\ha\ $<$ 0.4, that have  [O~I]:\ha\ $\ge$ 0.1, indicating these low-ionization lines tend to track one another.

We can summarize these results as follows: If a SNR candidate has a [S~II]:\ha\ $\ge$ 0.4, there are usually other indicators of shock heating that support its identification as a SNR.  Of the 160 SNRs satisfying the [S~II]:\ha\ $\ge$ 0.4 criterion, 154 have [S~III]:[S~II] $<$ 0.2, 113 have [O~I]:\ha\ $\ge$ 0.1, and 137 have \ha\ FWHM $\ge$ 3 \AA.  However, we also note that many of the SNR candidates which fail the  [S~II]:\ha\  criterion would pass as SNRs based on relatively weak [S~III] emission and/or evidence of velocity broadened lines.  Of the 69 SNR candidates that fail the [S~II]:\ha\ criterion, 48 have [S~III]:[S~II] $<$ 0.2, and 34 have \ha\ FWHM $\ge$ 3 \AA.  One way to interpret this would be to argue that a number of the objects observed to be below the [S~II]:\ha\ threshold are indeed likely to be SNRs.

By comparison, in the 188 object H~II region sample, 181 have [S~II]:\ha\ $<$ 0.4, so one would expect little contamination with the SNR sample.  However, 97 of these objects have [S~III]:[S~II] $<$ 0.2, 11 have [O~I]:\ha\ $\ge$ 0.1,  and 11 have \ha\ FWHM $\ge$ 3 \AA, all more similar to the SNR sample.  Of the seven \hii\ objects that have [S~II]:\ha\ $\ge$ 0.4, six have [S~III]:[S~II] $<$ 0.2, three have [O~I]:\ha\ $\ge$ 0.1 and three have \ha\ FWHM $\ge$ 3 \AA, and thus have SNR-like characteristics. Since these faint H~II regions were selected ``by eye'' looking only at the MUSE \ha\ data, indications are that a small amount of confusion in the sample is likely at the lowest surface brightnesses sampled.

\begin{figure}
\plotone{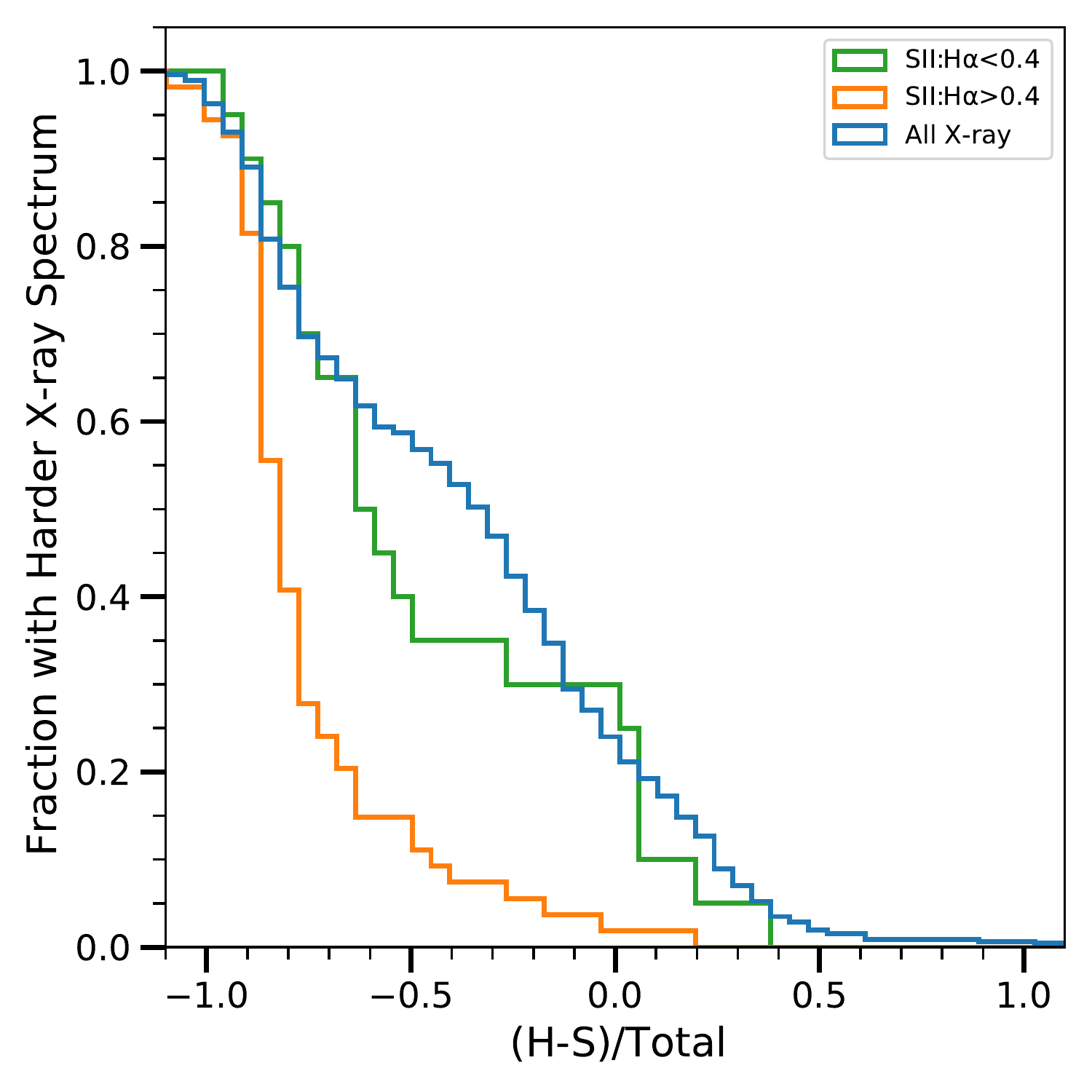}
\figcaption{The fraction of X-ray sources with hardness ratios $(M-S)/T$ greater than a given value for all the sources in the X-ray sample, for sources spatially coincident with SNR candidates with [S~II:]\ha\ $\ge$ 0.4 and for those with ratios less than 0.4.     \label{fig:xray}}
\end{figure}

{
Another largely independent indicator that an emission nebula is a SNR is evidence of X-ray emission, since H~II regions (though not necessarily the X-ray binaries that can be found in them) are relatively faint in X-rays.  
A  total of 108 of the SNR candidates in our SNR catalog lie within 1\arcsec\ of an X-ray source identified by \cite{long14} in M83.  Shifting the positions of the sources in random directions and recalculating the number of spatial coincidences suggests that no more than 10 of these coincidences are expected by chance. Of these, 74 were in the regions observed with MUSE, and 54 of those have [S~II]:\ha\ $\ge$ 0.4.  }

{One way to identify SNRs is through the hardness of the X-ray spectra. As discussed by \cite{long14}, SNRs are soft X-ray sources compared to  the other types of sources, such as X-ray binaries and background AGN,  that are typically seen in X-ray surveys of nearby galaxies, including M83. \cite{long14} characterized the X-ray spectra of M83 sources in terms  a hardness ratio of the form  $(M-S)/T$ where S is the counts observed between 0.35-1.1 keV, M between 1.1-2.6 keV, and T between 0.35-8 keV.  This ratio is expected to vary roughly from -1 for a source with  counts only in the S band, to +1 for sources with  counts only in the M band.} 

{As shown in Fig.\ \ref{fig:xray}, the 54 objects with coincident X-ray sources and [S~II]:\ha\ $\ge$ 0.4 are systematically softer in X-rays than the general X-ray source population in M83, which strengthens the case that these sources are actually SNRs.  
On the other hand, the distribution of hardness ratios of the SNR candidates with  [S~II]:\ha\  $<$ 0.4 is very different, which suggests, but does not prove (given that the sample contains only 20 objects), that most of these objects are not SNRs.  
To make this somewhat more quantitative, 
a Kolmogorov-Smirnov test of the hypothesis that the hardness ratios of SNRs with [S~II]:\ha\  $\ge$ 0.4 are drawn from the same population as the entire X-ray sample is disproven with a probability of $1.3~\times ~10^{-10}$, but the same test results in a value of 0.28 for the objects with [S~II]:\ha\ $<$ 0.4. } 

{In principle, radio emission also provides a straightforward way to distinguish H~II regions and SNRs.  Indeed, most Galactic SNRs were first identified as extended, non-thermal X-ray sources.  In the case of M83, \cite{russell20} identified 270 individual radio sources in M83 (outside of the complex nuclear region) using ATCA.  Of these, 62 lie within 1\arcsec\ of the current sample of SNR candidates in M83, whereas seven would have been expected by chance.   There are 38 of these objects with MUSE spectra, and 26, or 68\% have [S~II]:\ha\ $\ge$ 0.4, which is fairly similar to the fraction that have high [S~II]:\ha\  ratios  in the entire MUSE sample.   \cite{russell20} hoped to use their radio data to identify SNRs in M83 but,  unfortunately, they found that the radio spectral indices derived from the their data were not accurate enough to separate thermal and non-thermal radio sources. Hence, the existing catalog is of limited utility in determining whether any particular SNR candidate is actually a SNR. Hopefully a radio survey that delivers reliable spectral indices will be conducted in the not too distant future.\footnote{{\cite{russell20} carried out a more sophisticated analysis of spatial coincidences than we have done here that accounted for the apparent sizes of the radio sources.  They identified 64 in unconfused regions based on the \cite{williams19} list of SNR candidates. To avoid future confusion and because of the limited utility of the present radio spectral data, we have chosen not to include radio coincidences in Table \ref{snr_master}.} } }

\section{Limitations of the \sii:\ha\ ratio criterion \label{sec:limitations} }

It is worth revisiting the history of using the [S~II]:\ha\ ratio as a primary shock diagnostic. 
Although  \cite{mathewson73} and other early researchers proposed the  strengths of \sii\ lines relative to \ha\ as an optical criterion for distinguishing SNRs from \hii\ regions, \citet{dodorico78} appears to have been the first to  
propose using the specific ratio of \sii:\ha\ = 0.4 as the observational dividing line between shock-heated and photoionized gas.  This criterion was then adopted by \cite{blair81} in their early study of SNRs in M31, and in many other studies going forward.  The expectation of enhanced \sii:\ha\ ratios was also vetted by many early shock model calculation grids, such as those of \cite{raymond79} and \cite{shull79}, although a specific dividing line of 0.4 was not called out.  Indeed, close inspection of these models, or more recent grids of shock models that cover a broad parameter space such as those of \cite{allen08}, show that there are indeed shock conditions for which [S~II]:\ha\ ratios below 0.4 can result.  The application of this criterion has been widely successful not so much because of the specific value of 0.4 but because observationally there has been a large gap between the typical low [S~II]:\ha\ ratios seen in photoionized gas and the enhanced ratios well in excess of 0.4 observed in many SNRs. In recent studies over the last decade, it has become clear that this convenient empirical diagnostic begins to break down as one assembles larger (and generally fainter) samples of SNRs, and in particular as one also compares with fainter and perhaps more typical \hii\ regions.  Below, we consider two different regimes where confusion in applying the [S~II]:\ha\ ratio criterion occurs.

\subsection{Are the Candidates that have \sii:\ha\ $<$ 0.4 likely to be SNRs?}

As shown in the left panel of Fig.\ \ref{fig:s2_ratio}, applying a ratio of 0.4 across the sample raises a number of questions. As discussed earlier, \cite{winkler17} identified a small number of SNR candidates whose GMOS spectra showed ratios below the 0.4 threshold (see their Fig.\ 6). In \cite{winkler17}, we argued that most of these nebulae should still be regarded as good candidates, 
either because a) there was other information that suggested they were SNRs, b) not all shocks have [S~II]:\ha\ ratios greater than 0.4, or c) because  uncertainties in the measurements made their actual ratios uncertain.  These objects were examined and determined to be good SNR candidates despite the low ratios.  Have we altered our opinion,  based on the new MUSE observations?   Our answer is no, for a combination of observational and theoretical reasons.  

There are numerous bright candidates for which the MUSE-derived ratio is below 0.4. Two significant but related factors that could contribute to this are a) contamination by coincident \hii\ emission and b) uncertainties in background subtraction.  It turns out that of the 20 candidates with high surface brightness that do not pass the [S~II]:\ha\ $ \ge 0.4$ criterion,   nearly all are in the nuclear region where both the above problems are most  extreme.  Our choice to take a background from the area of lowest surface brightness within 5\arcsec\ of the SNR candidate is conservative in the sense that we avoid over-subtracting the background, but means we are sensitive to emission that lies very close to the SNR, especially as these sources, many of which were discovered on the basis of the [Fe II] emission,  are faint in the optical.  Even though the MUSE observations were taken under good (0\farcs7) seeing, a definitive assessment of the nature of these objects really requires the angular resolution  of HST.

\begin{figure}
\plotone{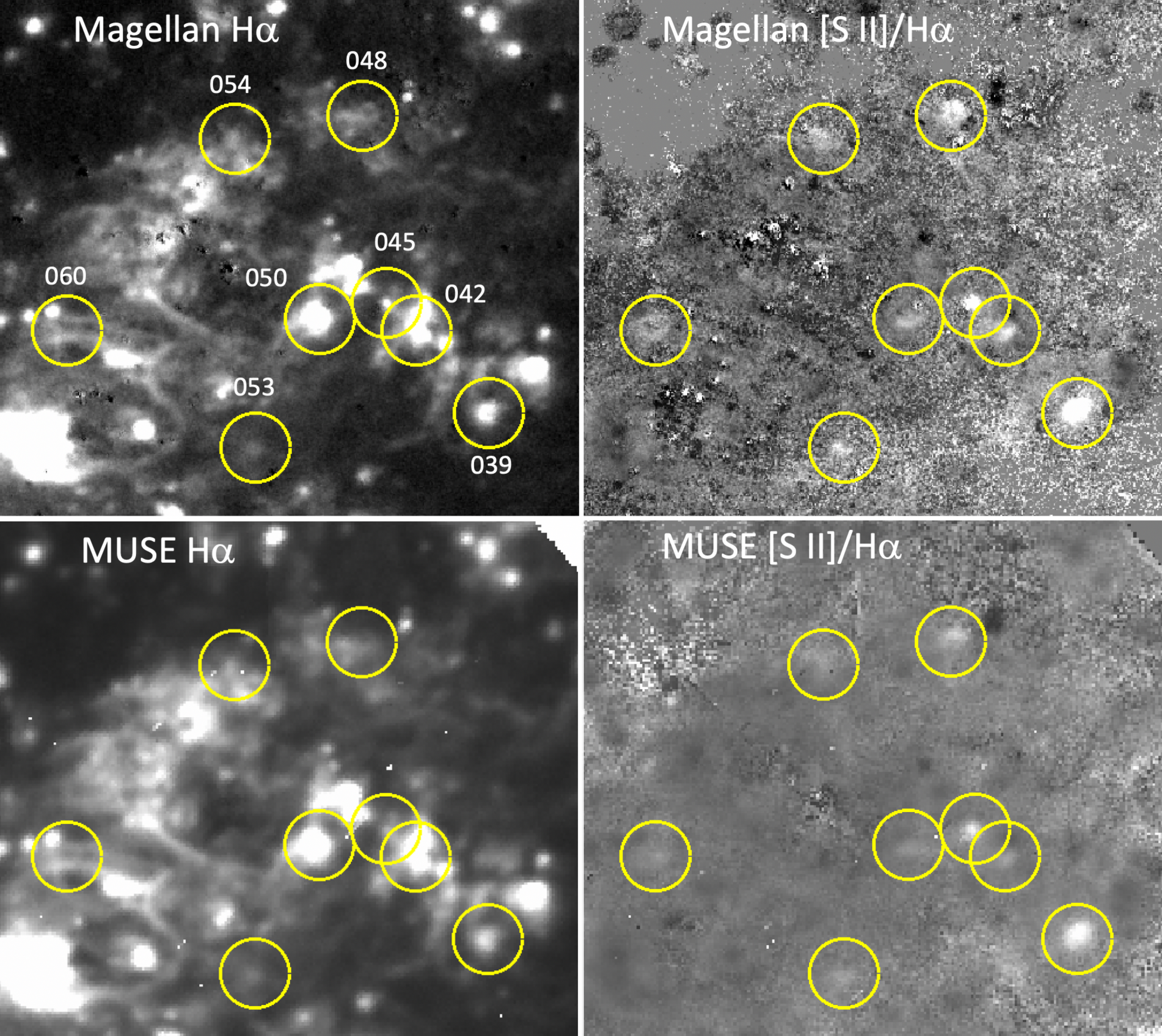}
\caption{A figure similar to Fig.\ \ref{fig:muse_new} showing a $\sim$35\arcsec\ region in the southwest spiral arm of M83.  The yellow circles indicate previous SNRs from \cite{blair12}, with numbers in the upper left panel indicating their identifications in that catalog. These objects are embedded in extended \ha\ emission at various levels, and their [S~II]:\ha\ ratios are only modestly enhanced in the Magellan ratio map.  However, most of these objects are even less enhanced in the MUSE ratio map, likely due to the somewhat lower spatial resolution (and hence additional contamination) of the MUSE data.  This effect also impacts the observed ratios in the extracted MUSE spectra for such objects.  The scaling on the ratio maps is from 0 (black) to 1.5 (white). \label{fig:muse_magellan}}
\end{figure}

Fig.\, \ref{fig:muse_magellan} shows a cluster of SNRs from \cite{blair12} embedded in the extended emission region in the spiral arm to the southwest of the M83 nucleus, as they appear in the Magellan imaging data and with MUSE.  The objects of interest are seen in the Magellan ratio map with modestly enhanced ratios relative to surrounding emission.  However, the MUSE ratio map shows less enhancement for a number of these objects despite the narrower MUSE bandpasses extracted, as discussed in Sec.\, \ref{sec:more}.  Possibly, the 0\farcs7 resolution of the MUSE data is just enough poorer than the 0\farcs4-0\farcs5 resolution of the Magellan IMACS data that smearing and \hii\ region contamination lower the observed ratios. Though the problem is not as severe as in the nuclear region, our belief is that these objects remain viable candidates, especially factoring in the Magellan results.

\begin{figure}
\plotone{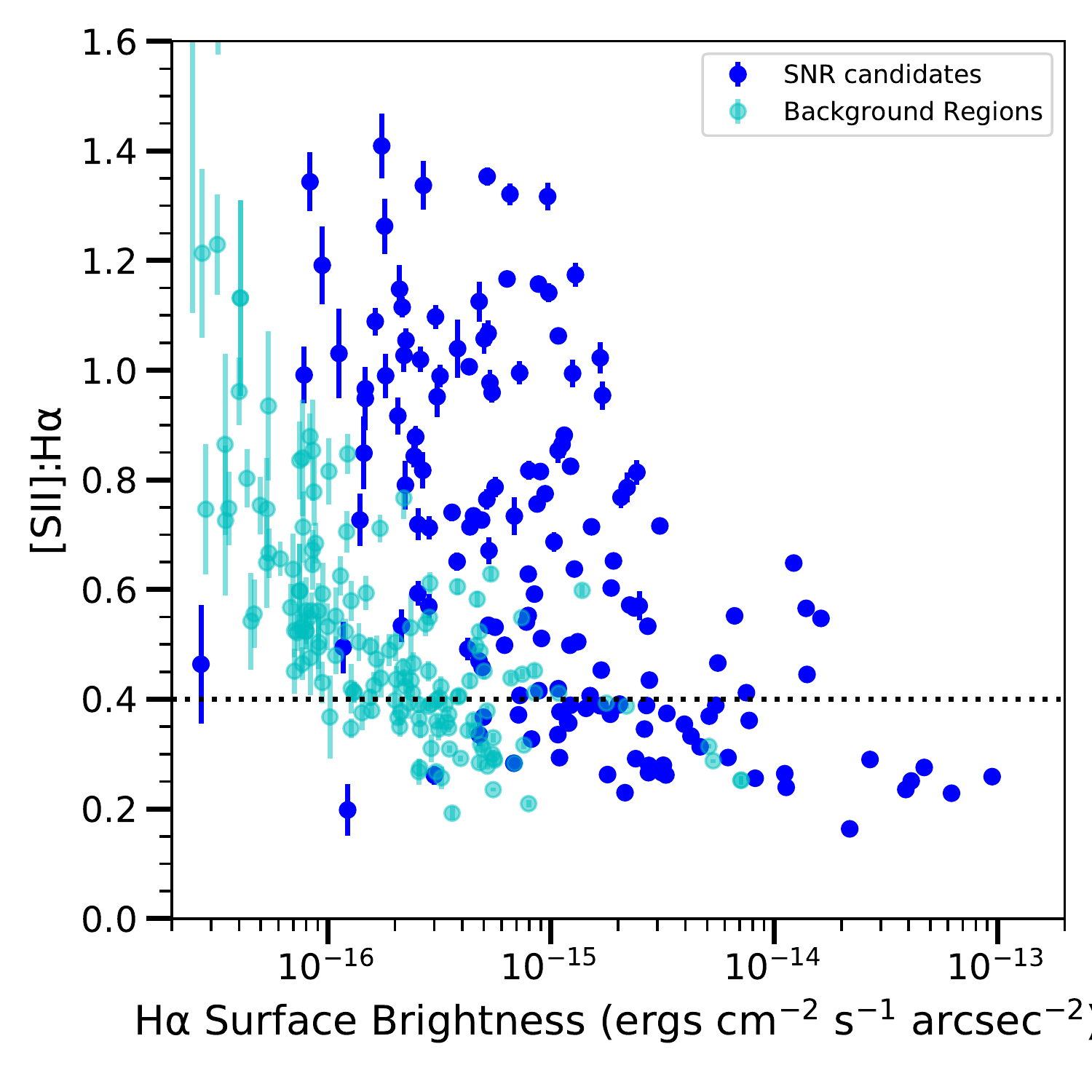}
\figcaption{The \sii:\ha\ ratios of  SNR candidates observed with MUSE compared to that of the spectra obtained from the background regions associated with the SNR candidates.  Emission nebulae with [S~II:\ha\ ratios greater than 0.4 are normally considered good SNR candidates.  However, at lower surface brightness levels, the ratio rises into the nominal SNR range, indicating that the normal value may not be applicable.  See text for details.    \label{fig:dig}}
\end{figure}

Furthermore, \cite{kopsacheili20} have carried out a re-examination of the optical line ratios that might serve to identify SNRs in nearby galaxies by comparing the theoretical line ratios produced by the shock models of \cite{allen08} to two sets of photoionization models of gas around starbursts created by \cite{kewley01} and \cite{levesque10}.  Both the shock and the photoionization models were carried out with Mappings \citep{sutherland93,allen08}, and therefore should be directly comparable.   \cite{kopsacheili20}  note that a large number of the shock models have [S~II]:\ha\ ratios less than 0.4, and  in particular that using 0.4 as a strict cutoff disfavors the identification of SNRs with slow shock velocities.\footnote{One caveat to this conclusion is the absence of shock models below 100 $\VEL$ in the \cite{allen08} model grid.}   If these shocks are realized in nature, then  one should continue to study the set of objects that have [S~II]:\ha\ ratios that are somewhat elevated compared to H~II regions to see whether there are other indicators that they actually are shocks.

We note in passing that  \cite{kopsacheili20} argue that various combinations of line ratios involving [O~I], [O~II], [O~III], [N~II] and [S~II] provide a more accurate way to identify all nebulae with shocks.   A full analysis of all of these possibilities for the MUSE spectra  is beyond the scope of this paper and would not be straightforward due to the large range in S/N of the various spectra.  As an example of the difficulties, we note that \cite{kopsacheili20} argue that the cleanest single line for identifying shocks is the [O~I]:\ha\ ratio; of the models they investigate, 97\% of shocks had a ratio greater than 0.017, but only 2.4\% of the starburst models satisfied this ratio.   The observational problem, of course, is that measuring [O~I] to a precision that is such a small percentage of \ha\ is difficult, which is why we used a value of 0.1 above as a secondary indicator that a nebula was shock dominated.  Also, while the statistics \cite{kopsacheili20} quote are correct in terms of what the models predict, the relative number of instances of each model in nature is unlikely to be flat.  In the case of the M83 sample,  113 of the 160 SNR candidates satisfying the [S~II]:\ha\ criterion of 0.4 also have [O~I]:\ha\ greater than 0.1, but only three of the other candidates has a measured [O~I]:\ha\  greater than 0.1.  By contrast,  for the H~II sample of 188 objects, 14 have [O~I];\ha\ greater than 0.1;  if the \cite{kopsacheili20}  actually reflected nature, these objects would be classified as SNRs.  

\subsection{Are faint objects with \sii:\ha\ $> 0.4$ likely to be SNRs? \label{sec:faint} }

Transitioning to the fainter end of the distribution, over the last decade new SNR surveys have pushed to lower surface brightness, and the gap in ratio between H~II regions and SNRs has closed to the point where the middle ground is muddled.  Thus, one needs to consider the possibility that some of the emission nebulae we are identifying as SNRs are really just bright patches in the Diffuse Ionized Gas, or DIG, that exists in the more distributed ISM.  Despite having very low surface brightness in \ha, the DIG competes with \hii\ regions in terms of the total \ha\ luminosity emitted by a galaxy because of its spatial extent.   Observationally \cite[see, e.g.][for a review]{haffner09}, the DIG has many of the same properties as  gas in SNRs, and specifically high [S~II]:\ha\ ratios that correlate inversely with low surface brightness \citep{galarza99}.   As measured by   \cite{bruna21}, the radially averaged surface brightness of the DIG in M83 is in the range \EXPU{2-2.5}{-16}{ergs \: cm^{-2} s^{-1} arcsec^{-2}}.

One way to assess how serious this problem might be is to compare the \sii:\ha\  ratios obtained from the background-subtracted spectra of the SNR candidates to that of the spectra obtained from the background regions themselves, which is shown in Fig.\ \ref{fig:dig}.   A number of trends are apparent:  (1) The \sii:\ha\ ratios of the background regions show a pronounced trend toward higher ratios at low surface brightness.  If these background regions were well-defined nebulae, which they are not,\footnote{Recall the difference between the background regions and the faint SNR candidates we have identified is that the SNR candidates were identified as coherent, identifiable emission regions whereas the background regions could be random locations within 5\arcsec\ of the nominal SNR coordinate. Hence, the background regions have systematically lower surface brightness and could well be representative of DIG in many cases.} many would be regarded as viable SNR candidates.  (2) A number of the faint SNR candidates  have very high \sii:\ha\ ratios compared the main locus of background points at the same surface brightness, which is encouraging.  However, the 34 SNR candidates with a surface brightness of   $<$\POW{-16}{ergs \: cm^{-2} s^{-1} arcsec^{-2}} are typically only a few times as bright as the background that is being subtracted from them.  We have checked to see if there are other properties that might distinguish the faint SNR candidates from the DIG, as represented by the background regions.  The measured widths of \ha, and the [O~III]:\hb, [O~I]:\ha, and [N~II]:\ha\ ratio distributions are fairly similar.  One might hope to distinguish low surface-brightness SNRs from the (photoionized) DIG based on the shape of the line profiles, but with a few exceptions, the low-surface-brightness candidates have lines that are unresolved at the MUSE resolution.
Hence, the identification of the faint objects as SNRs depends fairly critically on their being identified as well-defined nebulae. For these reasons, even though the faint objects have the spectroscopic characteristics of SNRs, it seems likely that some fraction of them could be misidentified.

\subsection{The Bottom Line}

As SNR surveys in nearby galaxies have pushed to lower surface brightnesses,
the oft-used observational discriminant of observed [S~II]/\ha $\ge$ 0.4 to
indicate shock-heated nebulae has
become less deterministic. While it remains true that bright photoionized
nebulae typically have low values of this ratio, at lower surface brightnesses
this is no longer the case and observed ratios can meet or exceed the normal
threshold to indicate shock heating. Of course, observational error in
determining the ratio also increases for the faintest nebulae and the proper
removal of any overlying  background emission also becomes more problematic.
Also, referring to grids of shock model calculations where many variables
come into play, it is clear that there are regions of parameter space where
shocked nebulae do not necessarily produce a ratio above 0.4, although
this remains true for a wide portion of the expected parameter space for shocks.

We and others, including \cite{kopsacheili20}, have investigated various secondary criteria that appear to be useful in specific cases to help confirm shock heating in optically-identified
candidates, but none of these provide a
silver bullet, especially for the faintest nebulae identified with the optical
criterion. For example:

\begin{itemize}
\item The presence of a coincident soft X-ray or non-thermal radio source is a
strong confirmation of shock heating and is even a primary diagnostic for SNRs
in our Milky Way. However, these emissions are typically  observable only for the brighter
objects in nearby galaxies, even with lengthy integrations.
\item The presence of additional shock heated line enhancements such as
[O~I]/\ha, [N~II]/\ha, [Fe~II]/\ha, or even the relatively new criterion used in this
paper, [S~III]/[S~II], can again help confirm in some individual cases, but are in many
ways similar to the [S~II]/\ha\ criterion in that they are variable with 
shock conditions and hence model-dependent.
\item The kinematic diagnostic should in principle be fairly deterministic;
photoionized gas should show thermal broadening which is only at the 10-20 
$\VEL$ level, while bulk motions from shocks should produce much broader line
profiles.  However, with only modest kinematic resolution of $\sim$100 $\VEL$
that is usually available, there is still the potential for indeterminate
results for slower velocity shocks. Obtaining even higher spectral resolution
data would be beneficial, but for exceedingly faint nebulae, this is a difficult task.
\end{itemize}

All of this is to say that the goal of achieving a complete survey of the SNR
population within a given external galaxy is very difficult to achieve. The veracity of our
SNR identifications is quite strong for the brighter objects, while the
identifications at lower surface brightnesses must be considered provisional
in many cases.

\section{Special Objects of Interest}

In a recent paper discussing some of the M83 SNR candidates that seemed to have peculiar morphology, \cite{soria20} concluded there was a pair of adjacent SNR candidates, B12-096 and B12-098, that were likely two lobes from a single microquasar. A hard stretch of the HST \ha\ image further shows a jet-like extension from B12-098 to the ESE.  The MUSE spectra of the bright emission from these two objects show similar spectra that confirm high \Sii:\ha\ ratios and moderately high \Sii\ densities for both lobes. \Nii\ $\lambda$6543 is comparable to or slightly weaker than \ha, and \Oiii\ $\lambda$5007 is comparable to or stronger than \ha.  In the extended red coverage provided by MUSE, \Siii\ $\lambda$9072 is clearly detected in both objects.
These spectra are thus consistent with shock heating of these lobes with shock velocities in excess of 100 $\VEL$.  This situation may be similar to that observed in the W50/SS433 microquasar system in our own galaxy \cite[][and references therein]{Dubner98}. This object may be a more evolved version of the first M83 microquasar that was detected just NE of the nuclear region \citep{soria14}.

Another object worthy of mention is B12-174a, an apparently very young SNR ($\le$100 yrs) that shows very high velocity ($\sim$5000 $\VEL$) emission features in its spectrum \citep{blair15}. B12-174a is a bright compact emission knot that appears in projection against the northern limb of a larger, lower surface brightness SNR candidate, B12-174.  The MUSE B12-174a spectrum is shown in Fig.\ \ref{fig:example} and indicates the presence of narrow lines in addition to the broad red emission feature that encompasses the entire \ha-\Nii-\Sii\ region. The MUSE spectrum of the larger shell, B12-174, confirms a shock-heated spectrum, but the flux from this larger SNR is too faint to account for the narrow emission lines seen at the B12-174a position.  

Since these narrow lines show somewhat elevated \Sii:\ha\ and \Nii:\ha, it would be interesting if they were intrinsic to the young SNR, perhaps indicating a shock precursor or other low velocity shock emission in addition to the broad component.  
However, these narrow lines were not seen by \cite{blair15}. We have reviewed the GMOS data, that were obtained with a 1\farcs25 $\times$ 6\arcsec\ slitlet in good seeing. 
Performing spatial and spectral crosscuts in the 2D GMOS spectra, we do not find evidence of narrow emission associated directly with the compact SNR, although variable narrow emission overlays the entire region.  We conclude that the narrow features in the MUSE spectrum are {most likely residual background emission that is along the line of sight to the object. There is little or no difference  in the shape of the broad red component in the MUSE and GMOS spectra of B12-174a, although given the quality of both spectra, the changes would have to have been fairly large to have been detected. No new broad features were detected in the extended red spectral coverage provided by MUSE.}

\begin{figure}
\plottwo{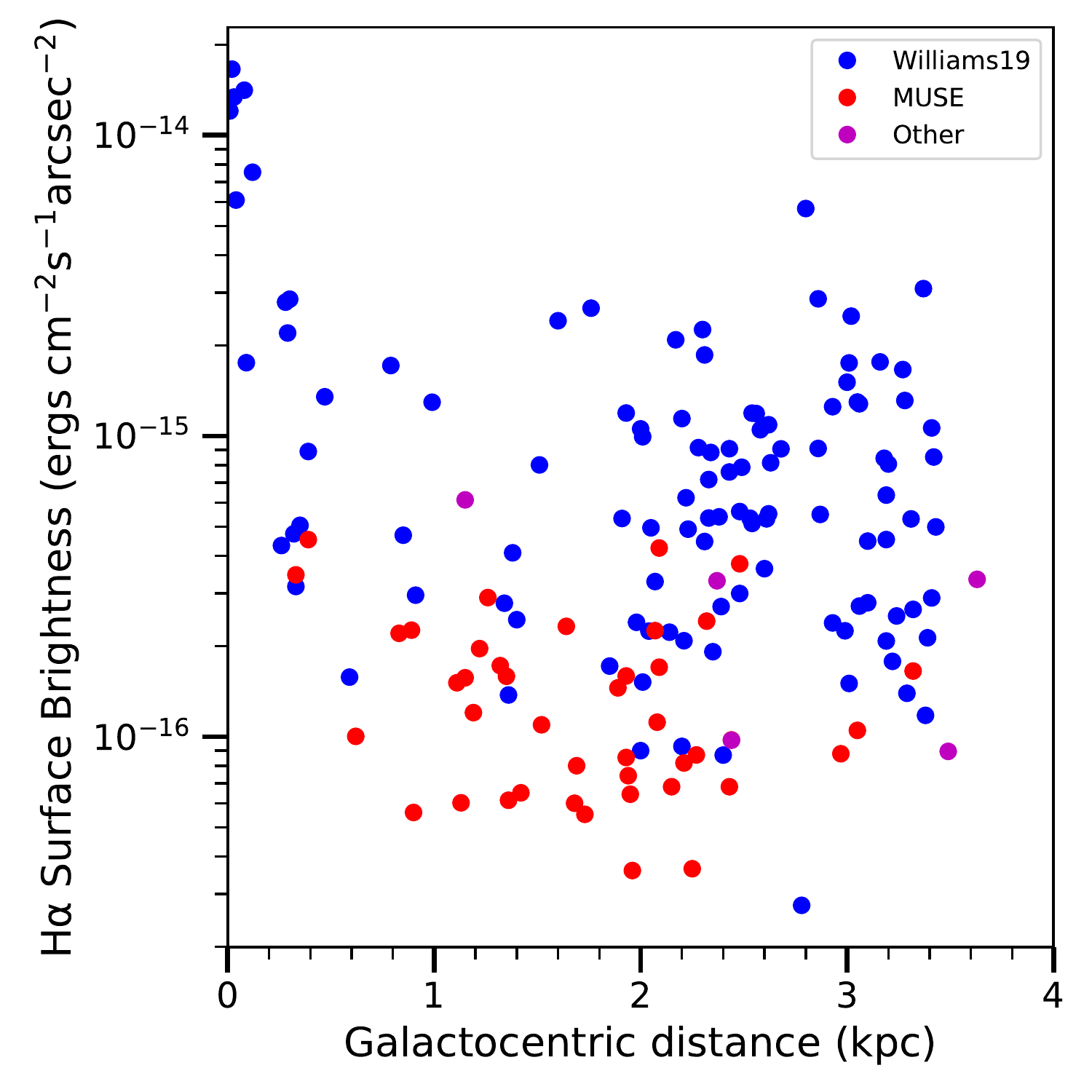}{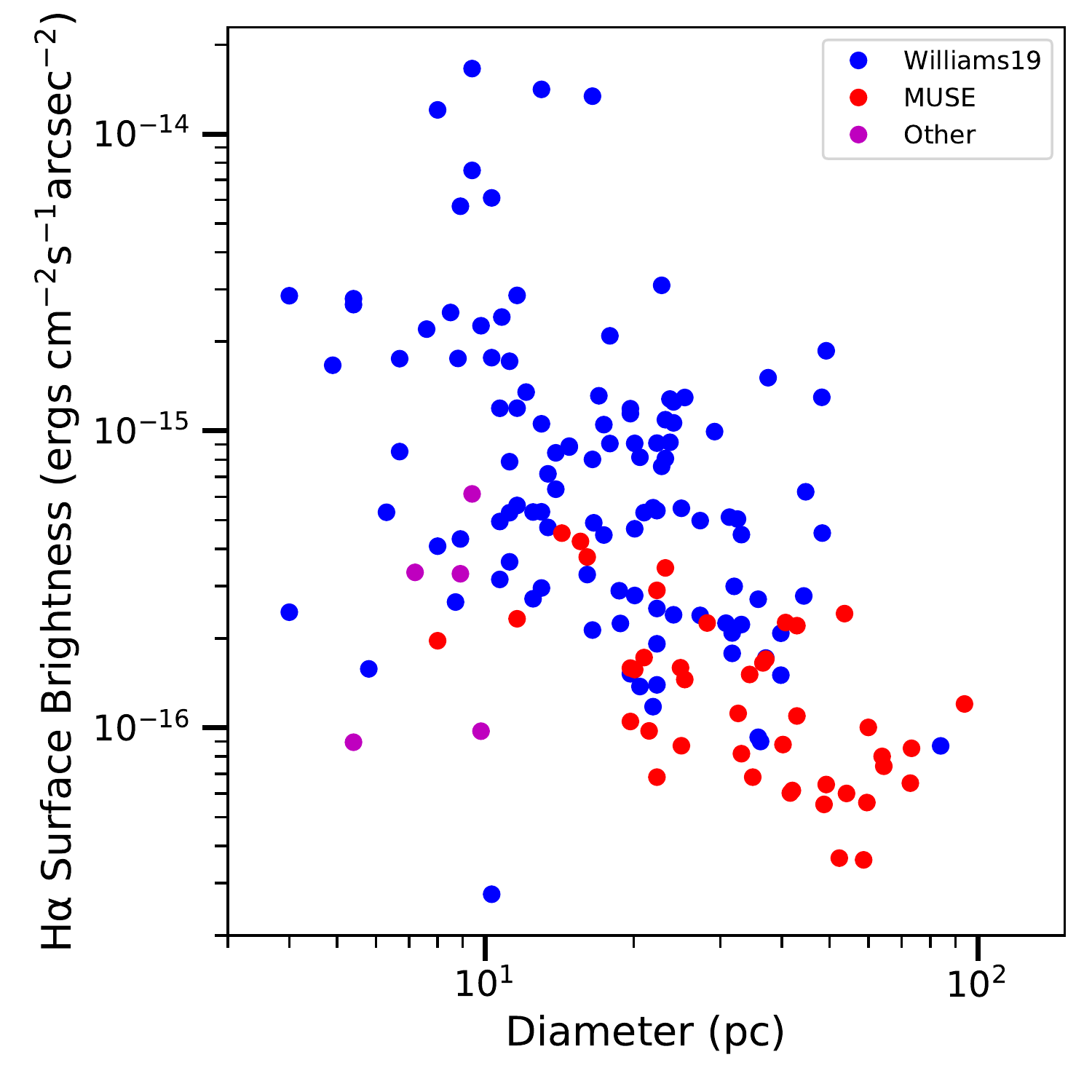}
\figcaption{  Left:  \ha\ surface brightness as a function of galactocentric distance for SNR candidates with [S~II]:\ha\ ratios greater than 0.4.  Right:  \ha\ surface brightness as a function object diameter for the same objects.  \label{fig:surf}}
\end{figure}

\section{Global Trends in the MUSE Spectra of SNRs in M83}

Given such a large and homogeneous sample of spectra, it is important to try to understand any systematic trends in the characteristics they display.  For this purpose, we consider only those SNR candidates that have [S~II]:\ha\ ratios greater than 0.4 and therefore are most likely to be SNRs.   Fig.\ \ref{fig:surf} shows the surface brightnesses of these objects as a function of galactocentric distance and object diameter.   Except in the nuclear region where the background is very high, there is no obvious trend of surface brightness with galactocentric distance.  This is consistent with the idea that the surface brightness of a SNR is dependent on local conditions, not global ones.  By contrast, there is a clear correlation of between surface brightness and SNR diameter.  This is at least partially a physical effect:  While smaller diameter objects with low surface brightness, and hence low total flux,  are unlikely to have been identified, larger diameter, high surface brightness objects would certainly have been found if present, and none are seen.  The most straightforward explanation for this is that high surface brightness suggests high density and the shocks of SNRs expanding into higher density evolve more rapidly than those expanding into lower density, and thus simply evolve beyond the radiative phase before reaching a large diameter.

\begin{figure}
\plottwo{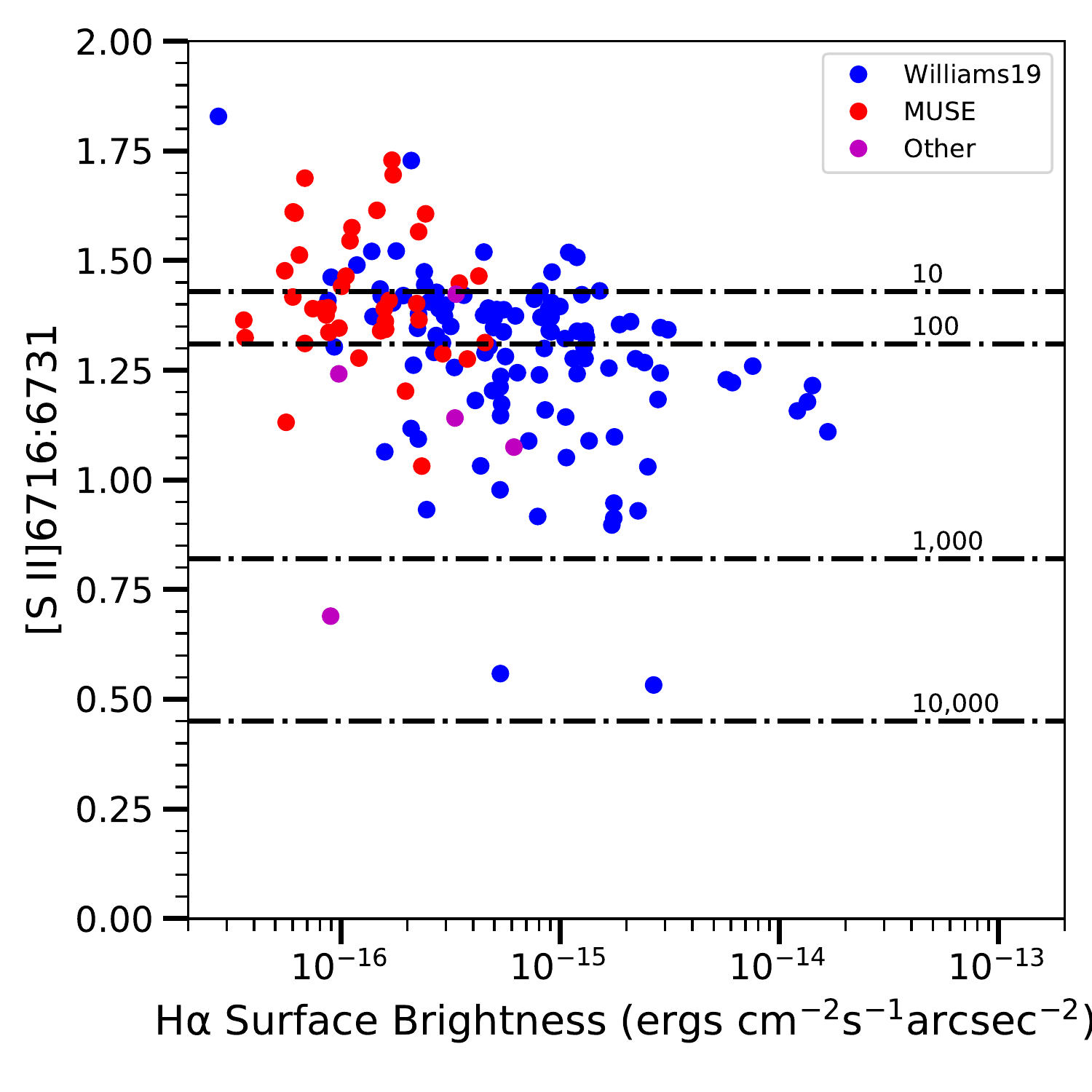}{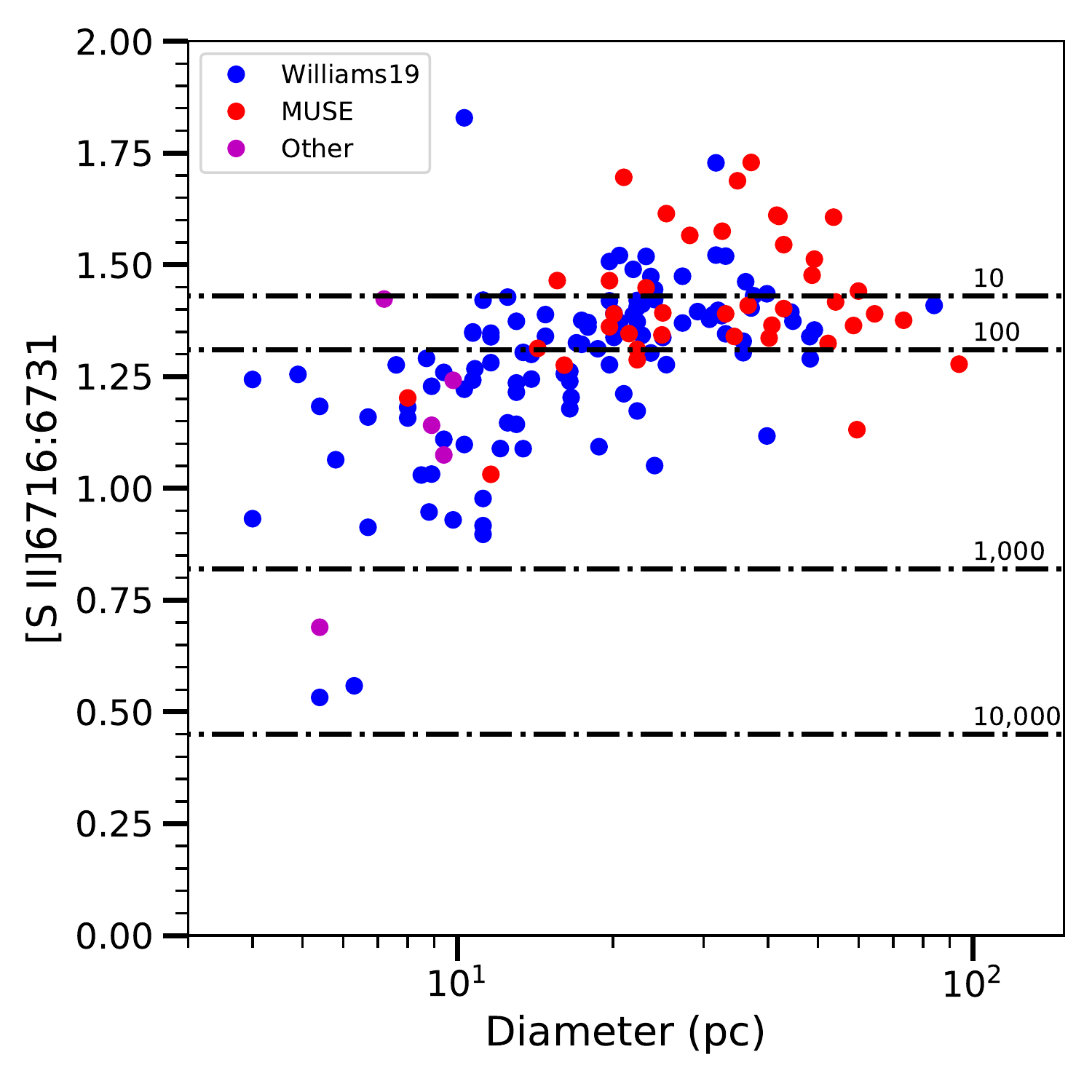}
\figcaption{  Left: The density-sensitive [S~II]6716:6731 ratio as a function  of \ha\ surface brightness for SNR candidates with [S~II]:\ha\ ratios greater than 0.4. Right:   [S~II]6716:6731 as a function of object diameter. The dashed lines show the expected ratios for for various electron densities \citep{osterbrock06}.   \label{fig:s2_den}}
\end{figure}

Support for the idea that smaller diameter objects tend to be expanding into denser media on average is provided by the density sensitive line ratio of [S~II]6716:6731.    Although the scatter is significant, Fig.\  \ref{fig:s2_den},  shows that objects with higher surface  brightness  (left panel) and smaller diameter (right panel) tend to have lower [S~II]6716:6731 ratios, which implies higher densities.  
The addition of the newly discovered MUSE low surface brightness sample tends to solidify these conclusions: although the observational errors increase for these faint objects, many are at or above the nominal low density limit of the ratio in the plot.

\begin{figure}
\plottwo{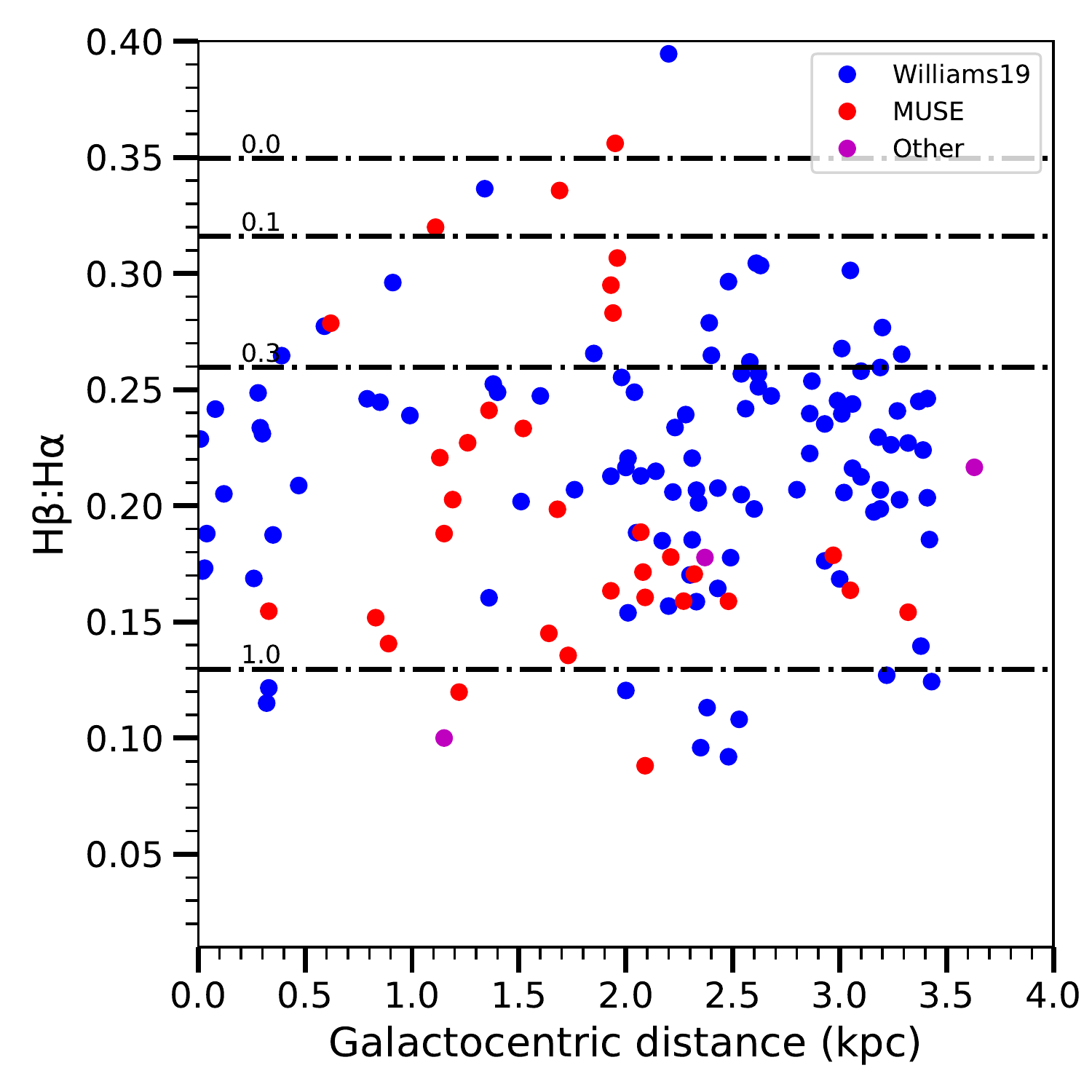}{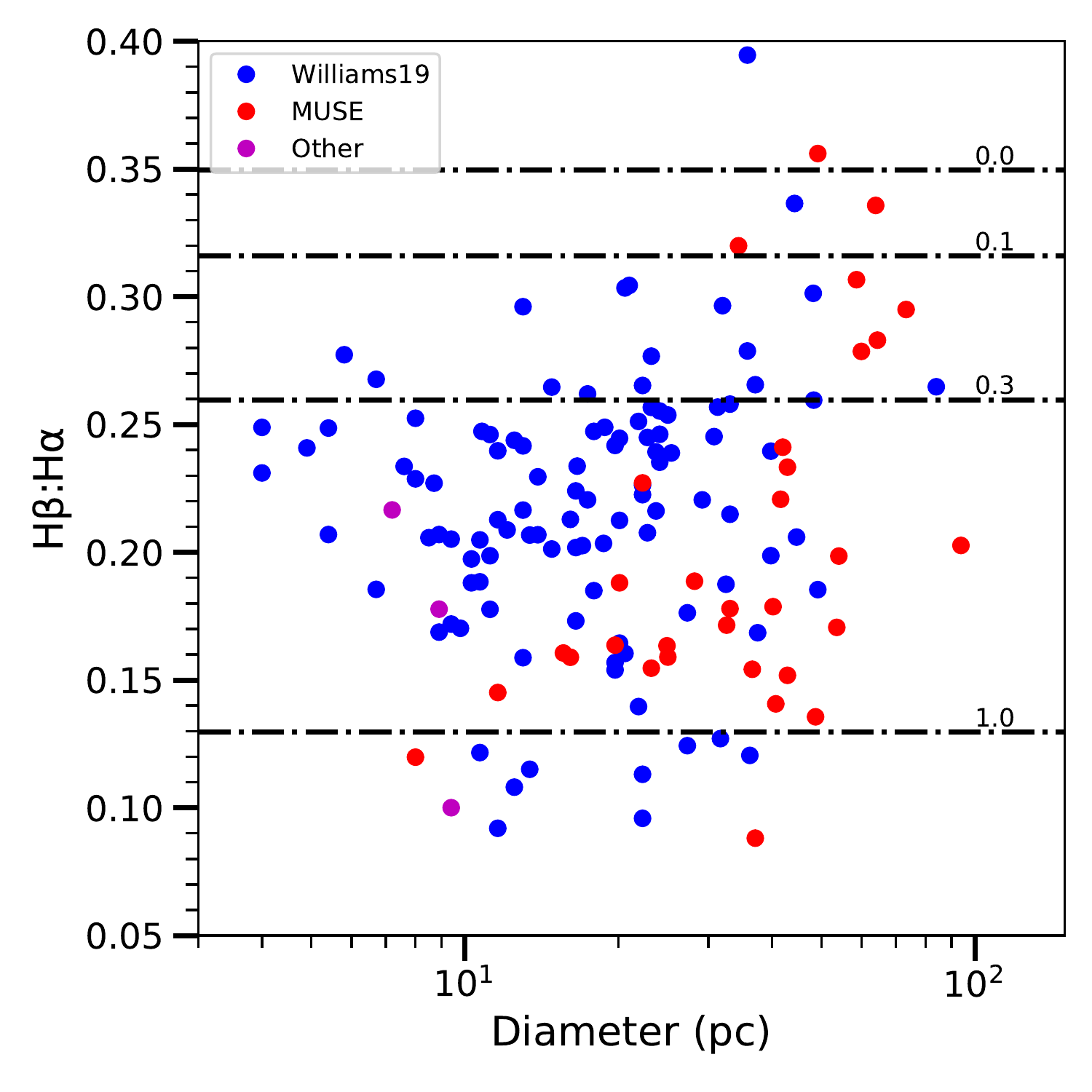}
\figcaption{  Left: \hb:\ha\ as a function of galactocentric distance for SNR candidates with [S~II]:\ha\ ratios greater than 0.4.  Right:   \hb:\ha\ as a function object diameter.  The dashed lines show the expected \hb:\ha\ ratios for various values of E(B-V) \citep{seaton79}. \label{fig:reddening}}
\end{figure}

Most of the objects in the SNR sample show evidence for a moderate amount of reddening as shown in Fig.\ \ref{fig:reddening}.   The galactic foreground extinction should be small \cite[E(B-V) = 0.059;][]{schlafly11} so the very low values seen for a few objects likely point to a small amount of observational error.  There are no obvious trends in the amount of reddening with either galactocentric distance or diameter, as expected since extinction effects should be dominated by very local conditions impacting the individual objects.  The range seen here is directly comparable to Fig.\ 7 of \cite{winkler17} from the GMOS spectroscopy. The newly identified MUSE objects span the entire range of extinction seen in the earlier sample, although  once again, the uncertainties are larger for these faint objects, especially in measuring the \hb\ line in the more reddened objects.


\begin{figure}
\gridline{
\fig{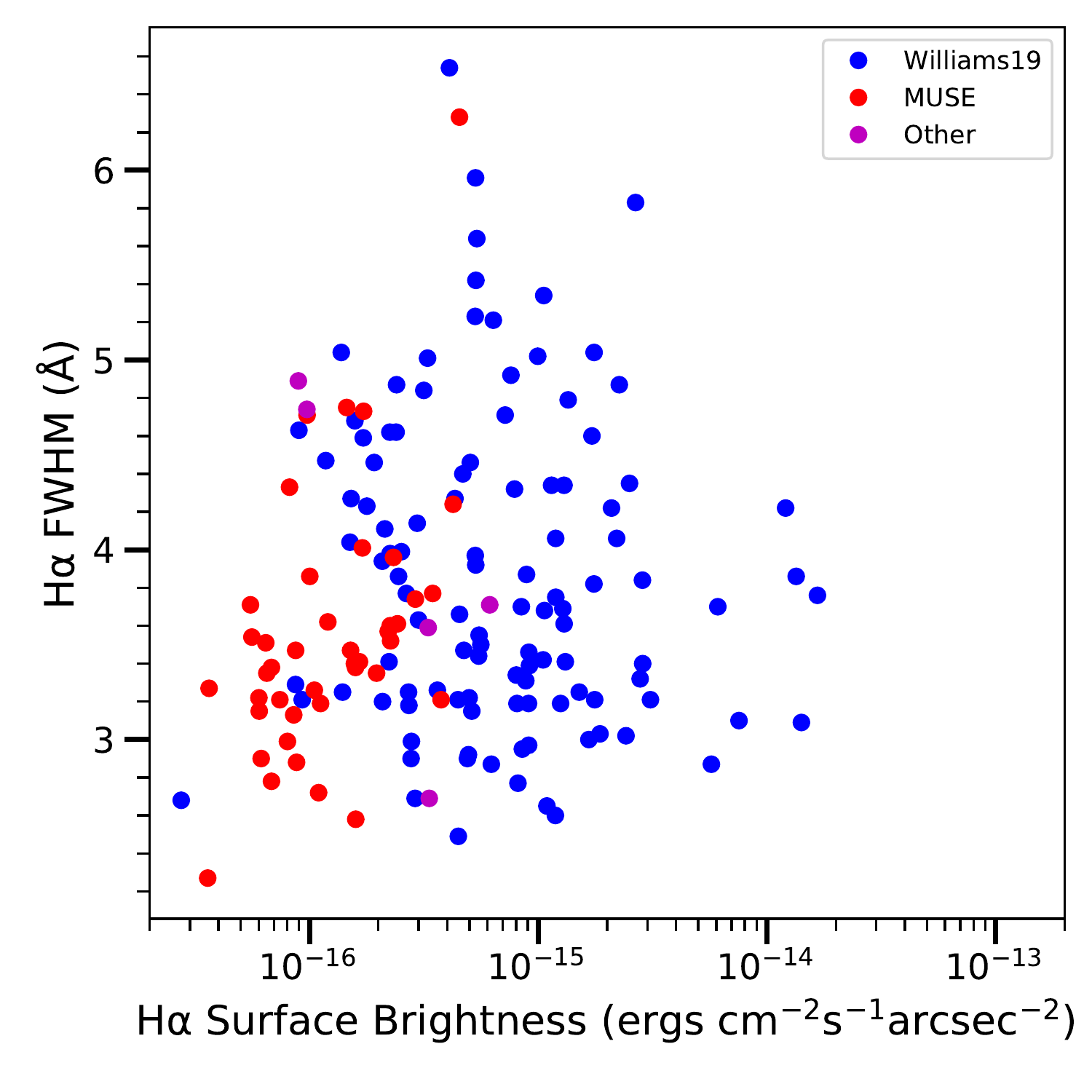}{0.3\textwidth}{(a)}
\fig{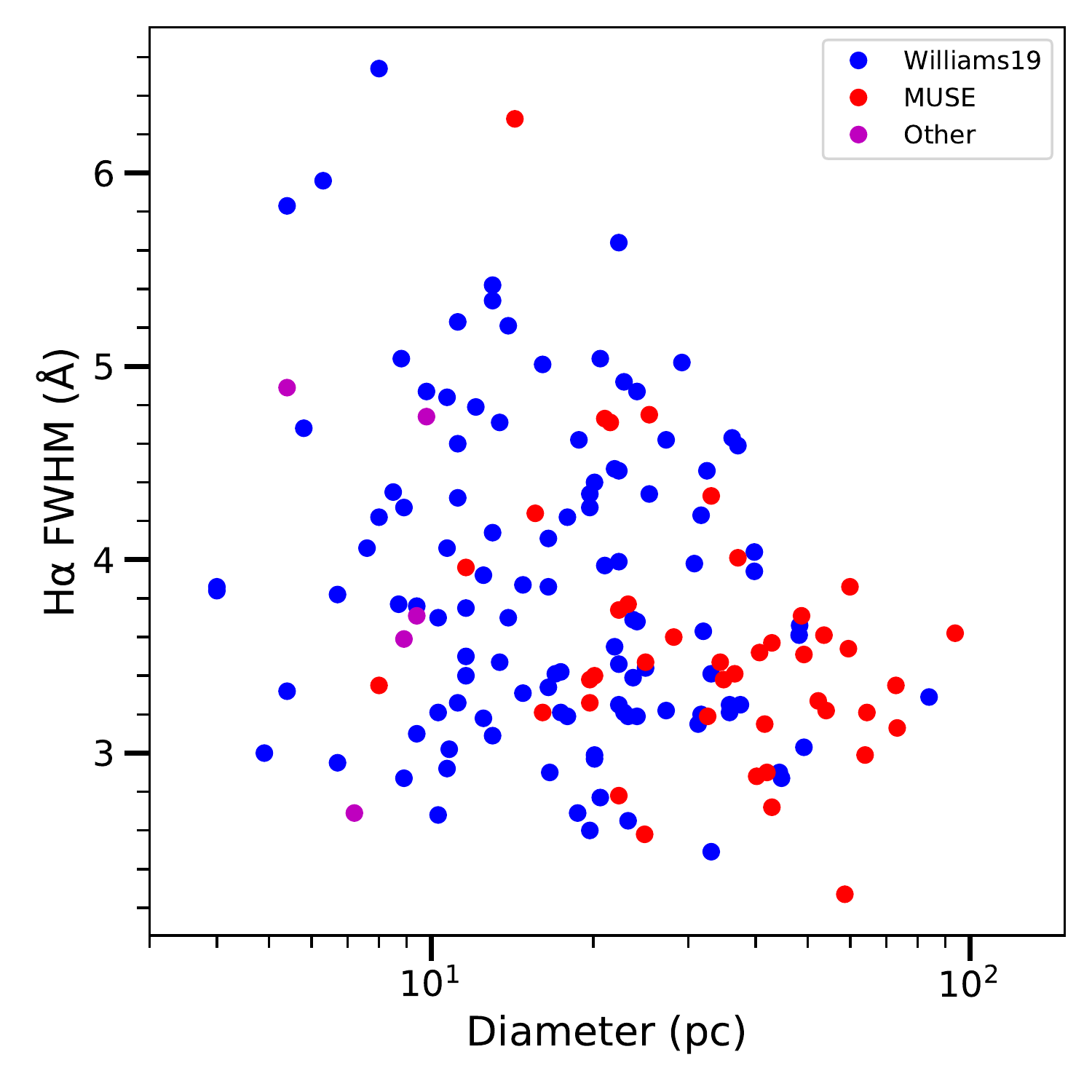}{0.3\textwidth}{(b)}
\fig{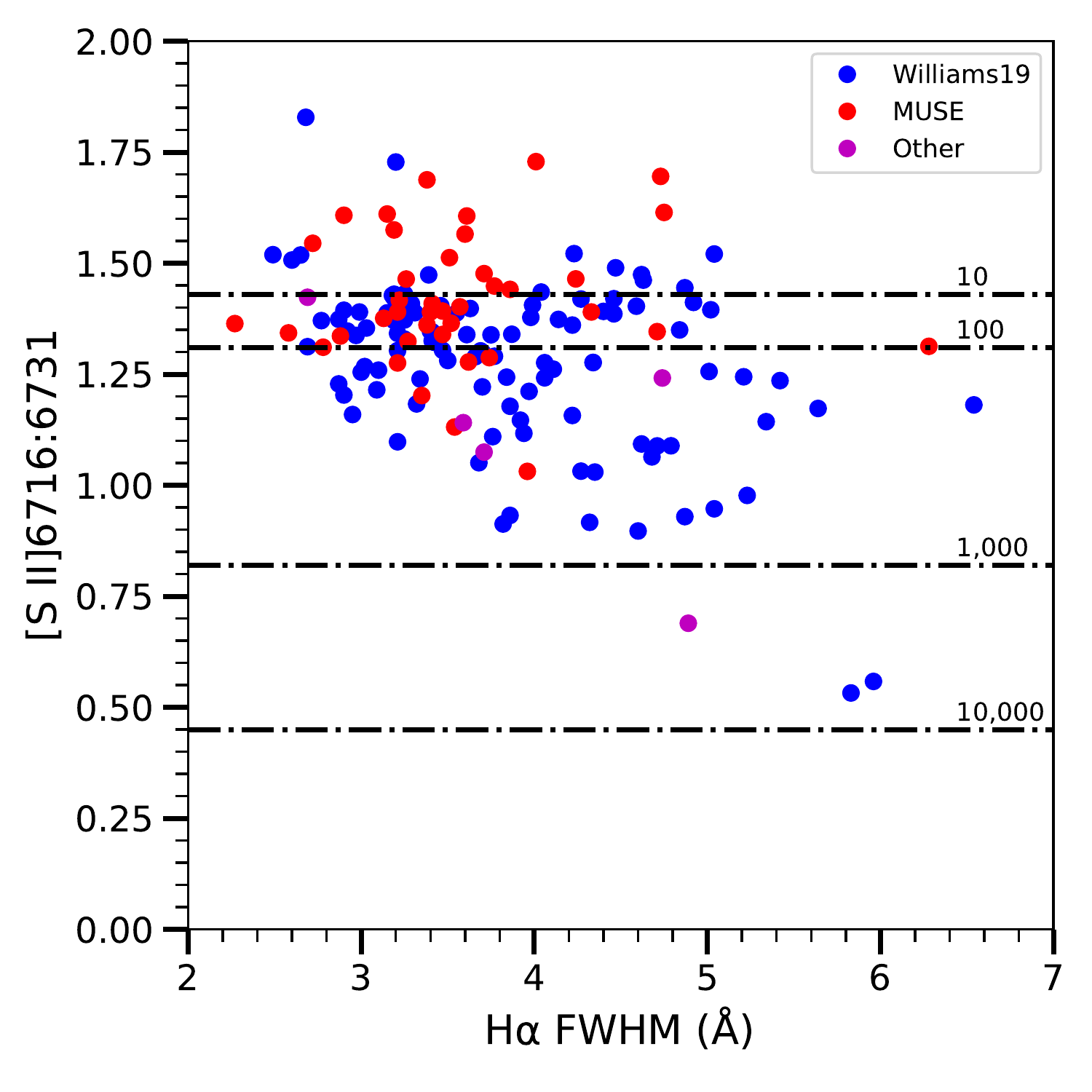}{0.3\textwidth}{(c)}
}
\figcaption{ (a) \ha\ FWHM as a function of \ha\ surface brightness  for SNR candidates with [S~II]:\ha\ ratios greater than 0.4.  (b) \ha\ FWHM as a function of object diameter.  (c) The density sensitive [S~II]6716:6731 ratio as a function of \ha\ FWHM. \label{fig:fwhm}}
\end{figure}

As shown in Fig.\ \ref{fig:fwhm},  correlations between FWHM of \ha\ as a function of surface brightness and diameter are weak.  That said, the range of FWHM for objects with diameters less than {20} pc is clearly larger than the range of FWHM for objects larger than this.  The FWHM in the spectra represent the dispersion in the bulk velocity of the shocked material in a SNR (since the thermal velocity of material behind the shock is small).  In principle, therefore, one could get a low FWHM with a high velocity shock, if all of the emission came from a single shocked cloud at an edge of SNR.  However, if the shocked material is distributed at various positions around the periphery, it is more likely that the FWHM correlates with the typical shock velocity.  Thus, the large dispersion in FWHM at smaller diameters is consistent with the interpretation that many of these SNRs have encountered higher density gas than others.  The objects at large diameters have all evolved to the point where the shock velocity is relatively low.   

While the scatter is large, objects that exhibit  a large FWHM in \ha, and hence higher shock velocities, tend to have  [S~II]6716:6731 ratios associated with higher density, as one would expect if SNe explode and eject (to an order of magnitude or so) the same amount of energy into the ISM in the form of shocks.  If the ISM is dense, the SNR will evolve to the limit of detectability relatively quickly and the energy will be radiated away on a relatively short period of time.  If the ISM is tenuous the SNR will become detectable later, when the velocity of the primary shock is lower.  SNRs expanding into dense media reach the radiative phase at smaller diameters, after which they begin to fade.   The optical emission from SNRs arises primarily from secondary shocks that are driven into the denser regions of the ISM, as is evidenced by the fact that the optical and X-ray appearance of resolved SNRs is usually quite different.  However, there is rough pressure equilibrium between the primary and secondary shocks, since the secondary shocks are driven by the pressure mismatch between the X-ray gas and the cool clouds.  Therefore, we expect that small diameter SNRs will tend to show higher densities and higher velocity widths than larger objects.

\section{Comparison to Models}

\begin{figure}
\plottwo{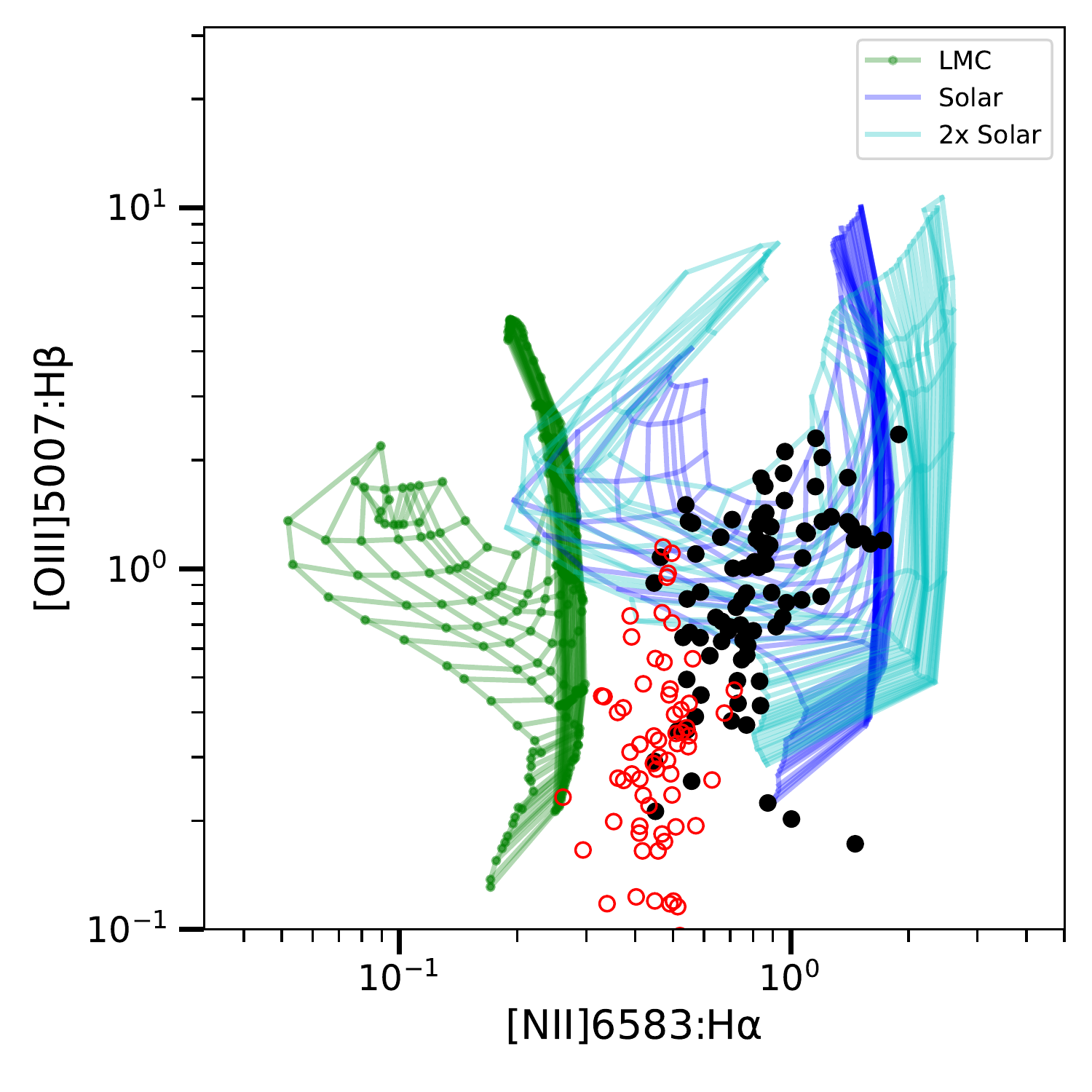}{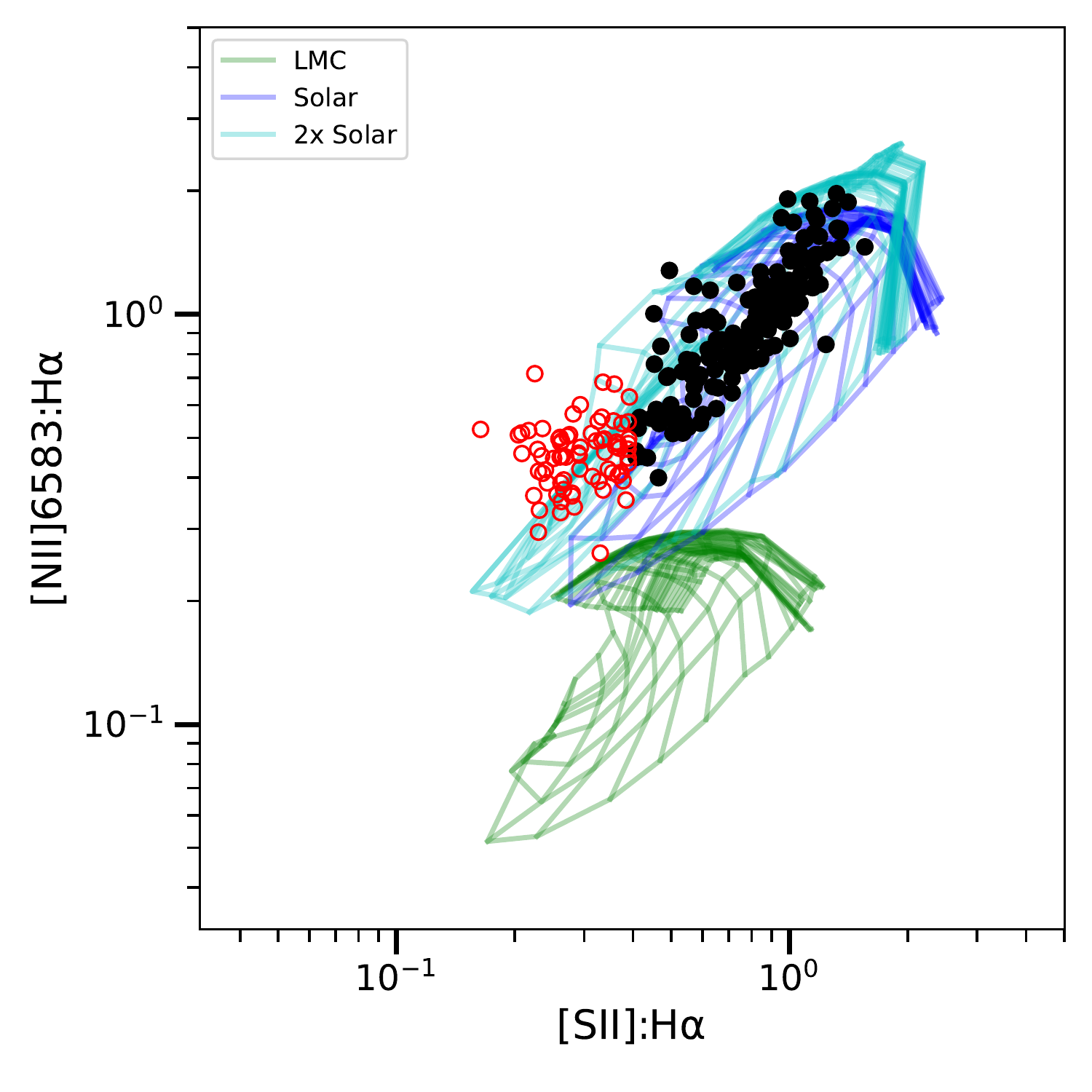}
\figcaption{Left: Observed [O~III] 5007:\hb\ ratio as function of the [N~II] 6583:\ha\ line ratio for {M83} SNRs and SNR candidates with spectra compared with model grids from \cite{allen08}. As discussed in the text, the green, blue and cyan meshes correspond to shock models with a range of shock velocities and pre-shock magnetic fields, and with metallicities corresponding to LMC (green), solar (blue) and twice solar (cyan), respectively.  Candidates that satisfy the [S~II]:\ha\ criterion are plotted in black; those that fail this test are plotted as open red circles.  Right: The same comparison for [N~II] 6583:\ha\ line ratios as a function of the [S~II]:\ha\ ratios.   \label{fig:models}}
\end{figure}

The emission seen from a typical SNR in M83 represents emission from a large number of shocks traversing material with a range of densities in local circumstellar (for small objects) and/or interstellar material.  Therefore, the line ratios that are observed are not expected to correspond exactly to those obtained from calculations based on a single shock velocity and density.  Nevertheless, comparisons to theoretical models have typically proven useful for characterizing the global spectra of extragalactic SNRs.  

Fig.\  \ref{fig:models} shows grids of the expected line ratios of [O~III]:\hb\ vs. [N~II]$\lambda$6583 and [N~II]$\lambda$6583:\ha\ vs. [S~II]:\ha\ from a series of shock models calculated by \cite{allen08} using the Mappings III code. The discrete points show the results from our spectra.  The model grids span a range of shock velocity from 100 to 1000 $\VEL$ and pre-shock magnetic field from 10$^{-4}$ to 10 $\mu$G, with different colors representing abundance sets that are appropriate for LMC, solar, and twice solar values.    All of the models are for a pre-shock density of 1 cm$^{-3}$ and all ignore precursor ionization of the ISM.\footnote{\cite{allen08} also calculated models allowing for precursor ionization.  The main difference qualitatively is that at least for solar and twice-solar abundances, the [N~II]$\lambda$6583:\ha\  ratios are lower and the [O~III]:\hb\  ratios are somewhat higher.  The [O~III]:\hb\  of objects that fail the [S~II]:\ha\ criterion still fall outside the range of the models.} 

A number of trends can be seen in Fig.\  \ref{fig:models}. First, none of the observations fall within the locus of the models created with sub-solar LMC abundances, in contrast to a similar comparison for SNRs in M33, which has abundances that are about 1/2 solar \citep{long18}.   Rather, nearly all of the observed data points fall within the region described by theoretical line ratios in the solar or twice solar [N~II]$\lambda$6583:\ha\ vs. [S~II]:\ha\ diagram, including those candidates which do not satisfy the normal [S~II]:\ha\ criterion. This is  expected since, according to  \cite[][]{bresolin16},  M83 has an oxygen metallicity of  9.0, or $2.2 \times$\,solar in the nuclear region, and 8.8 ($1.5 \times$\,solar) at 0.4 R25. 
However, nearly all the objects that do not satisfy the [S~II]:\ha\ criterion have very low [O~III]:\hb\ ratios compared to models in the twice-solar grid.  In both of these diagnostic plots, there is a general correlation of the observed spectra between the ratio of  [O~III]:\hb\ vs. [S~II]:\ha\ as well as [N~II]:\ha\ vs. [S~II]:\ha.)

As we have noted earlier, \cite{kopsacheili20} have pointed out that many of the  \cite{allen08}  shock models have [S~II]:\ha\ ratios of less than 0.4. This is evident in Fig.\  \ref{fig:models}. On the other hand, most of the SNR candidates we have observed in M83 with MUSE that have low [S~II]:\ha\ ratios have O~III]:\hb\ ratios that fall outside the locus of the theoretical models.  To the extent that one has faith in the models, this would argue that the objects with  low [S~II]:\ha\ ratios are not SNRs.  

An alternative possibility is that these SNRs have emission that is dominated by lower shock velocities than available in the comparison model grid. The \cite{allen08} models only start at 100 $\VEL$ and go up. At lower shock velocities, it could be that oxygen never gets ionized to $\rm O^{++}$, and hence, little [O~III] emission is present.  This is consistent with the discussion from \cite{blair14} where many of the SNRs in M83 become bright and radiative at small diameters due to a high density, high pressure ISM. By the time the SNRs expand to intermediate and larger diameters, the shock velocity decreases, and the ionization goes down.  This explanation is also consistent with the trends noted above (e.g., lower surface brightness and lower densities for larger diameter SNRs).

\section{Summary and Conclusions}

Historically, SNRs in external galaxies have been identified using narrow-band interference-filter imaging to identify candidates, and with moderate resolution spectroscopy to confirm that the candidate objects have high [S~II]:\ha\ ratios compared to \hii\ regions.  In M83, after the corrections noted in Sec.~3, the \citep{williams19} list contained  278 SNR and candidates, of which 118 had spectra \citep{winkler17}.  Of these, 103 had [S~II]:\ha\ ratios that exceeded 0.4, the dividing line most often utilized to declare a SNR candidate as a bona fide SNR in the optical.

Here, we have used a mosaic of MUSE data for M83, as  assembled by \cite{bruna21},  along with our earlier HST and Magellan images of M83 to update and study the SNRs in M83.  Our new catalog of SNRs contains {366} objects, of which {81} have not been reported {previously, including 44 SNR candidates} obtained by inspection of narrow-band images extracted from the MUSE data cube.

There are 229  of these catalog objects contained within the region observed with MUSE, 170 of which have never previously been observed spectroscopically, making a substantial step forward in characterizing the overall SNR population.  We have extracted spectra for all  candidates, and our analysis shows that 160 have  [S~II]:\ha\  ratios greater than 0.4.    Of the objects that satisfy this criterion, 137 also show lines that are broadened beyond that seen in a sample of H~II regions, and 150 have a ratio of [S~III]:[S~II] less than 0.2, as expected for shocked gas.  Many are also associated with Chandra-detected X-ray sources \citep{long14}.  By all pre-existing standards these objects are SNRs.  If we add SNR candidates outside the MUSE region that have spectroscopically determined [S~II]:\ha\  ratios greater than 0.4, then 211 SNRs have been confirmed in M83 out of the 288 that have been observed spectroscopically  with either GMOS or MUSE.  Most of the others that have spectra with measured [S~II]:\ha\ ratios well in excess of the value of 0.1 usually associated with bright H~II regions, but their true nature remains more uncertain.

These results contrast with MUSE spectra extracted for a set of 188 nebulae, nominally H~II regions, which were selected as random faint isolated patches of \ha\ emission.
These nebulae were intended to span the range of surface brightnesses in the SNR candidate sample to make a valid comparison.  Overall, the  [S~II]:\ha\  ratios for this sample were low, as expected; only seven of these randomly selected nebulae turned out to have a [S~II]:\ha\ ratio greater than 0.4, only 14 had a measured FWHM of \ha\ greater than 3\AA, and only 14 showed believable [O~I] emission.  

Limiting ourselves to the most confident set of SNR candidates for which the  [S~II]:\ha\ ratio was observed to be greater than 0.4, we find a number of trends:

\begin{itemize}
\item
The mean surface brightness in \ha\ declines with diameter, but (excluding the nuclear region)  not with galactocentric distance. In the nucleus, bright background limits sensitivity and so the objects we can identify are systematically brighter there.
\item
The observed density-sensitive  [S~II] 6716:6731 ratios increase with SNR diameter, indicating that the optical shocks seen in small diameter objects are, on average, encountering denser material than those in large diameter objects.
\item
SNRs with smaller diameters ($<${10} pc as measured on HST images)  have \ha\ line widths that span a larger range than those which have larger diameters.  This is consistent with diverse CSM/ISM environments surrounding the younger objects. 
\item 
SNRs with higher \ha\ line widths tend to be dominated by shocks with [S~II]6716:6731 ratios that indicate higher densities. This is expected if the postshock pressure is proportional to the shock velocity squared and the temperature of the [S~II] emission zone is always at a temperature of about 10,000 K.
\end{itemize}

These trends are largely consistent with the idea that small diameter objects can be small either because they are ``younger'' in an evolutionary sense or because they are encountering denser gas and hence  evolving more rapidly.  The large diameter objects must have been the result of SNe that exploded in regions with more tenuous gas, so that that they could expand to a larger size before becoming radiative and then fading away.

Compared to SNRs identified in other galaxies and to models of shocks in SNRs, the spectra of SNRs in M83 show relatively low [O~III]:\hb\ ratios. This may point to a population of SNRs in M83 with relatively low shock velocities that cannot ionize the gas to $\rm O^{++}$.

Finally, as a coda to this study, we note that relatively wide field, high spectral and spatial resolution  integral field spectrographs on large telescopes and with wide wavelength coverage, like MUSE, are certain to be increasingly important for  future optical  studies of SNRs in nearby galaxies.  This is in part because IFU spectrographs allow a cleaner separation of the relevant emission lines than is possible with the  narrow-band interference-filter imaging that has historically been used to identify SNR candidates in nearby galaxies.  The broad wavelength coverage makes available additional lines, in this case [S~III]$\lambda$9069, to help discriminate shock-heated from photoionized gas.  And most importantly, high spectral resolution allows one to identify SNRs kinematically, which is required to distinguish shock-heated from photoionized gas in fainter objects, since the line ratios seen from  the DIG  resemble closely those expected from shock-heated material.   The fundamental limitation, of course, is that to learn much about the global evolutionary properties of SNRs in external galaxies, one also needs to measure their diameters, which beyond the Local Group requires high spatial resolution imaging such as can be obtained with HST.

\begin{acknowledgements}
Based primarily on observations collected at the European Southern Observatory under ESO programmes 096.B-0057(A), 0101.B-0727(A), 097.B-0899(B) and 097.B-0640. KSL acknowledges partial support from he Space Telescope Science Institute, which is operated by AURA, Inc., under NASA contract NAS 5-26555. 
PFW acknowledges support from the NSF through grant AST-1714281.  WPB acknowledges partial support from the JHU Center for Astrophysical Sciences during the time of this work. AA acknowledges the support of the Swedish Research Council, Vetenskapsr\aa{}det, and the Swedish National Space Agency (SNSA).  Finally, we appreciate the advice of an anonymous referee whose careful reading of the manuscript allowed us to improve the final version of the manuscript.
\end{acknowledgements}


%

\vspace{5mm}

\facilities{VLT/MUSE, HST/WFC3, Magellan/IMACS}


\software{astropy \citep{astropy} }



\pagebreak

\bibliographystyle{aasjournal}

\bibliography{m83}





\begin{deluxetable}{lrrrrccccl}
\tabletypesize{\scriptsize}
\decimals
\tablecaption{Supernova Remnants and Candidates in M83}
\tablehead{
\colhead{Source name$^a$} &
\colhead{RA} &
\colhead{Dec} &
\colhead{D} &
\colhead{R} &
\colhead{X-ray$^b$} &
\colhead{Origin} &
\colhead{Spectra} &
\colhead{[S II]:H$\alpha>$0.4} &
\colhead{Comments$^c$}
\\
\colhead{} &
\colhead{(J2000)} &
\colhead{(J2000)} &
\colhead{(pc)} &
\colhead{(kpc)} &
\colhead{} &
\colhead{} &
\colhead{} &
\colhead{} &
\colhead{}
}
\startdata
B12-001 & 204.16663 & -29.85976 & 34 & 6.5 & -- & W19 & GMOS & GMOS & - \\ 
B12-002 & 204.16811 & -29.85181 & 11 & 6.5 & -- & W19 & none & no & - \\ 
B12-003 & 204.17040 & -29.85491 & 12 & 6.3 & X019 & W19 & GMOS & GMOS & - \\ 
B12-004 & 204.17292 & -29.87107 & 20 & 5.9 & -- & W19 & none & no & - \\ 
B12-005 & 204.17325 & -29.83230 & 14 & 6.8 & -- & W19 & GMOS & GMOS & - \\ 
B12-006 & 204.17638 & -29.87149 & 24 & 5.7 & -- & W19 & none & no & - \\ 
B12-007 & 204.17801 & -29.87637 & 12 & 5.5 & -- & W19 & none & no & - \\ 
L22-001 & 204.18136 & -29.85193 & 16 & 5.5 & X029 & New & none & no & - \\ 
B12-008 & 204.18208 & -29.84608 & 21 & 5.6 & -- & W19 & none & no & - \\ 
L22-002 & 204.18240 & -29.85858 & 64 & 5.3 & -- & New & none & no & - \\ 
B12-009 & 204.18258 & -29.86978 & 28 & 5.2 & -- & W19 & none & no & - \\ 
B12-010 & 204.18600 & -29.84275 & 38 & 5.5 & -- & W19 & GMOS & GMOS & - \\ 
B12-011 & 204.18880 & -29.88549 & 34 & 4.9 & -- & W19 & none & no & - \\ 
B12-012 & 204.19026 & -29.87258 & 81 & 4.6 & -- & W19 & GMOS & GMOS & - \\ 
B12-013 & 204.19137 & -29.89288 & 27 & 4.9 & -- & W19 & none & no & - \\ 
B12-014 & 204.19340 & -29.89509 & 39 & 4.9 & -- & W19 & GMOS & GMOS & - \\ 
B12-015 & 204.19559 & -29.77824 & 69 & 8.8 & -- & W19 & none & no & - \\ 
B12-016 & 204.19637 & -29.92542 & 82 & 6.3 & -- & W19 & none & no & - \\ 
B12-017 & 204.19658 & -29.89761 & 29 & 4.8 & -- & W19 & none & no & - \\ 
B12-018 & 204.19676 & -29.89359 & 38 & 4.6 & -- & W19 & none & no & - \\ 
B12-019 & 204.19708 & -29.81771 & 15 & 6.0 & -- & W19 & none & no & - \\ 
B12-020 & 204.19928 & -29.85504 & 33 & 4.2 & -- & W19 & GMOS & GMOS & - \\ 
B12-021 & 204.19972 & -29.86278 & 49 & 4.0 & -- & W19 & GMOS & no & - \\ 
B12-022 & 204.20045 & -29.85941 & 34 & 4.0 & -- & W19 & GMOS & GMOS & - \\ 
B12-023 & 204.20123 & -29.87908 & 20 & 3.9 & X046 & W19 & GMOS & GMOS & - \\ 
B12-024 & 204.20188 & -29.86172 & 43 & 3.9 & -- & W19 & none & no & - \\ 
B12-025 & 204.20245 & -29.86814 & 52 & 3.8 & -- & W19 & GMOS & GMOS & - \\ 
L22-003 & 204.20312 & -29.87479 & 10 & 3.7 & X048 & New & none & no & - \\ 
B12-026 & 204.20414 & -29.88169 & 5 & 3.8 & -- & W19 & GMOS & GMOS & - \\ 
B12-027 & 204.20469 & -29.87358 & 26 & 3.6 & -- & W19 & none & no & - \\ 
B12-028 & 204.20569 & -29.88887 & 39 & 3.9 & -- & W19 & GMOS & no & - \\ 
B12-029 & 204.20641 & -29.86030 & 97 & 3.5 & -- & W19 & MUSE & no & - \\ 
B12-030 & 204.20669 & -29.88491 & 79 & 3.7 & -- & W19 & none & no & - \\ 
B12-031 & 204.20676 & -29.88713 & 36 & 3.8 & -- & W19 & GMOS & GMOS & - \\ 
B12-032 & 204.20680 & -29.84290 & 23 & 4.1 & -- & W19 & none & no & - \\ 
B12-033 & 204.20700 & -29.90115 & 23 & 4.4 & -- & W19 & GMOS & GMOS & - \\ 
B12-034 & 204.20716 & -29.84921 & 46 & 3.8 & -- & W19 & GMOS & GMOS & - \\ 
B12-036 & 204.20754 & -29.87138 & 4 & 3.4 & X053 & W19 & GMOS & GMOS & - \\ 
B12-035 & 204.20755 & -29.88564 & 31 & 3.7 & -- & W19 & GMOS & GMOS & - \\ 
B14-03 & 204.20882 & -29.87880 & 14 & 3.4 & -- & W19 & MUSE & no & [FeII] \\ 
\enddata
\tablenotetext{a}{Source names beginning with B12, B14, and D10 are from \cite{blair12}, \cite{blair14} and \cite{dopita10}, respectively.
Previoulsy unpublished candidates begin with L22.}
\tablenotetext{b}{X-ray sources in the \cite{long14} list within 1\arcsec\ of a SNR candidate}
\tablenotetext{c}{Objects with the comment [FeII] were first identified on the basis of detectable [FeII]1.67$\mu$ emission.}
\tablecomments{This table with 366 rows is available in its entirety in machine-readable form}
\label{snr_master}
\end{deluxetable}

\pagebreak
\begin{deluxetable}{rrrrrrrrrrr}
\tabletypesize{\scriptsize}
\decimals
\tablecaption{MUSE spectra of M83 Supernova Remnant Candidates}
\tablehead{
\colhead{Source} &
\colhead{H$\alpha$ SB$^a$} &
\colhead{H$\beta^b$} &
\colhead{[OIII]5007$^b$} &
\colhead{[OI]6300$^b$} &
\colhead{H$\alpha$} &
\colhead{[NII]6584$^b$} &
\colhead{[SII]6716$^b$} &
\colhead{[SII]6731$^b$} &
\colhead{[SIII]9069$^b$} &
\colhead{H$\alpha$ FWHM$^c$}
}
\startdata
B12-029 & 65.5$\pm$0.6 & 74.6$\pm$2.6 & 11.6$\pm$5.7 & 7.5$\pm$3.0 & 300 & 101.9$\pm$2.2 & 51.3$\pm$1.4 & 33.9$\pm$1.3 & 3.7$\pm$1.3 & 2.82$\pm$0.03 \\ 
B14-03 & 293.5$\pm$1.2 & 37.4$\pm$1.1 & 16.0$\pm$0.7 & 6.2$\pm$0.7 & 300 & 153.9$\pm$1.1 & 47.5$\pm$0.6 & 35.8$\pm$0.6 & 50.7$\pm$0.6 & 2.58$\pm$0.01 \\ 
B12-038 & 28.9$\pm$0.5 & 61.0$\pm$8.3 & 0.0$\pm$9.7 & 26.2$\pm$4.1 & 300 & 153.6$\pm$4.6 & 95.1$\pm$3.9 & 72.5$\pm$3.7 & 0.0$\pm$6.0 & 2.69$\pm$0.05 \\ 
B12-039 & 309.2$\pm$1.7 & 73.5$\pm$1.1 & 47.2$\pm$0.8 & 25.0$\pm$0.6 & 300 & 224.0$\pm$1.6 & 120.8$\pm$0.6 & 90.0$\pm$0.6 & 9.3$\pm$0.6 & 3.21$\pm$0.02 \\ 
B12-040 & 80.7$\pm$0.5 & 83.0$\pm$3.6 & 57.1$\pm$3.2 & 18.6$\pm$2.1 & 300 & 162.3$\pm$1.6 & 99.3$\pm$1.7 & 69.4$\pm$1.6 & 0.0$\pm$1.7 & 3.19$\pm$0.02 \\ 
B12-041 & 85.2$\pm$1.5 & 55.6$\pm$3.5 & 76.2$\pm$2.9 & 27.5$\pm$2.8 & 300 & 170.7$\pm$4.8 & 83.3$\pm$2.3 & 71.9$\pm$2.2 & 18.7$\pm$2.5 & 2.95$\pm$0.05 \\ 
B12-042 & 506.2$\pm$3.7 & 61.0$\pm$0.5 & 71.6$\pm$0.5 & 6.0$\pm$0.3 & 300 & 142.9$\pm$1.9 & 63.4$\pm$0.6 & 48.4$\pm$0.6 & 16.3$\pm$0.4 & 2.72$\pm$0.02 \\ 
B12-043 & 49.9$\pm$0.6 & 37.3$\pm$4.8 & 9.9$\pm$5.2 & 9.9$\pm$2.9 & 300 & 126.5$\pm$3.2 & 84.7$\pm$2.4 & 61.8$\pm$2.3 & 0.0$\pm$4.8 & 3.22$\pm$0.04 \\ 
B12-045 & 176.4$\pm$3.4 & 59.2$\pm$2.0 & 123.0$\pm$1.6 & 44.3$\pm$1.2 & 300 & 291.5$\pm$5.7 & 113.2$\pm$1.5 & 103.1$\pm$1.5 & 15.2$\pm$1.4 & 3.21$\pm$0.05 \\ 
B12-047 & 131.3$\pm$2.0 & 60.8$\pm$2.3 & 41.6$\pm$2.0 & 29.3$\pm$1.6 & 300 & 197.1$\pm$4.2 & 105.4$\pm$1.4 & 79.5$\pm$1.3 & 7.2$\pm$2.0 & 3.41$\pm$0.05 \\ 
B12-048 & 129.8$\pm$2.5 & 90.4$\pm$2.3 & 212.8$\pm$2.9 & 10.0$\pm$1.6 & 300 & 159.2$\pm$5.2 & 86.2$\pm$2.3 & 64.4$\pm$2.2 & 16.6$\pm$2.0 & 3.61$\pm$0.07 \\ 
B12-049 & 166.4$\pm$1.2 & 72.2$\pm$1.8 & 54.5$\pm$2.3 & 11.2$\pm$1.4 & 300 & 145.7$\pm$2.0 & 67.2$\pm$1.3 & 53.6$\pm$1.2 & 12.7$\pm$1.1 & 3.00$\pm$0.02 \\ 
B12-050 & 638.5$\pm$3.0 & 58.5$\pm$0.5 & 31.4$\pm$0.6 & 5.7$\pm$0.3 & 300 & 143.7$\pm$1.3 & 50.3$\pm$0.6 & 37.4$\pm$0.5 & 16.0$\pm$0.3 & 2.71$\pm$0.01 \\ 
B14-08 & 28.9$\pm$0.6 & 0.0$\pm$5.4 & 0.0$\pm$5.1 & 19.1$\pm$7.6 & 300 & 143.2$\pm$5.6 & 47.5$\pm$3.6 & 42.4$\pm$3.6 & 106.5$\pm$5.8 & 2.35$\pm$0.05 \\ 
B14-09 & 1168.5$\pm$2.6 & 46.8$\pm$0.3 & 8.1$\pm$0.3 & 3.7$\pm$0.2 & 300 & 126.4$\pm$0.6 & 40.7$\pm$0.3 & 32.1$\pm$0.3 & 25.9$\pm$0.2 & 2.63$\pm$0.01 \\ 
B12-053 & 110.2$\pm$0.9 & 56.9$\pm$2.2 & 19.6$\pm$2.6 & 11.8$\pm$1.3 & 300 & 115.7$\pm$2.1 & 65.9$\pm$1.2 & 46.0$\pm$1.2 & 0.0$\pm$1.2 & 2.77$\pm$0.02 \\ 
B12-054 & 182.6$\pm$0.9 & 70.4$\pm$1.1 & 19.2$\pm$1.0 & 10.0$\pm$1.3 & 300 & 124.7$\pm$1.2 & 68.9$\pm$0.6 & 46.8$\pm$0.6 & 8.3$\pm$1.0 & 2.77$\pm$0.01 \\ 
B14-10 & 2.8$\pm$0.3 & 0.0$\pm$45.7 & 0.0$\pm$42.4 & 281.8$\pm$88.8 & 300 & 115.5$\pm$29.9 & 95.0$\pm$22.2 & 51.9$\pm$21.3 & 0.0$\pm$33.3 & 2.68$\pm$0.35 \\ 
B12-056 & 21.3$\pm$0.6 & 67.2$\pm$13.5 & 209.1$\pm$12.6 & 54.7$\pm$7.1 & 300 & 394.0$\pm$8.8 & 172.7$\pm$6.1 & 136.9$\pm$5.8 & 0.0$\pm$5.5 & 4.11$\pm$0.09 \\ 
B12-058 & 36.2$\pm$0.6 & 59.6$\pm$4.4 & 77.6$\pm$5.1 & 42.7$\pm$3.0 & 300 & 254.0$\pm$4.4 & 136.7$\pm$2.6 & 96.2$\pm$2.4 & 12.1$\pm$4.5 & 3.26$\pm$0.05 \\ 
B12-057 & 17.8$\pm$0.5 & 38.1$\pm$8.2 & 137.6$\pm$11.5 & 66.6$\pm$8.4 & 300 & 364.1$\pm$8.1 & 185.9$\pm$7.8 & 122.1$\pm$7.2 & 0.0$\pm$9.6 & 4.23$\pm$0.09 \\ 
B12-060 & 136.2$\pm$0.7 & 67.8$\pm$2.7 & 13.2$\pm$1.8 & 7.6$\pm$1.3 & 300 & 120.2$\pm$1.3 & 65.9$\pm$1.0 & 42.4$\pm$1.0 & 9.0$\pm$1.7 & 2.93$\pm$0.02 \\ 
B12-061 & 109.1$\pm$0.5 & 77.0$\pm$1.3 & 15.4$\pm$1.3 & 10.5$\pm$0.9 & 300 & 135.8$\pm$1.2 & 76.2$\pm$0.7 & 50.2$\pm$0.7 & 7.7$\pm$1.2 & 2.65$\pm$0.01 \\ 
L22-005 & 30.7$\pm$0.7 & 129.4$\pm$12.5 & 30.6$\pm$10.1 & 32.5$\pm$13.6 & 300 & 68.3$\pm$5.5 & 27.6$\pm$4.5 & 69.1$\pm$5.2 & 22.4$\pm$5.8 & 2.38$\pm$0.06 \\ 
B12-062 & 11.8$\pm$0.5 & 41.9$\pm$18.4 & 243.3$\pm$21.7 & 40.5$\pm$15.7 & 300 & 309.3$\pm$12.7 & 194.7$\pm$13.1 & 130.7$\pm$12.1 & 0.0$\pm$10.6 & 4.47$\pm$0.16 \\ 
B12-063 & 8.7$\pm$0.2 & 79.4$\pm$13.2 & 215.9$\pm$15.4 & 39.9$\pm$16.6 & 300 & 326.8$\pm$8.2 & 160.5$\pm$8.1 & 113.9$\pm$7.5 & 0.0$\pm$11.6 & 3.29$\pm$0.08 \\ 
B12-064 & 90.8$\pm$0.6 & 49.3$\pm$3.6 & 10.5$\pm$2.4 & 7.8$\pm$2.2 & 300 & 164.1$\pm$1.6 & 72.2$\pm$1.5 & 53.9$\pm$1.4 & 17.7$\pm$2.5 & 2.97$\pm$0.02 \\ 
B12-065 & 175.1$\pm$3.8 & 80.3$\pm$2.9 & 104.0$\pm$2.6 & 70.9$\pm$1.8 & 300 & 232.0$\pm$6.2 & 65.1$\pm$1.9 & 71.3$\pm$1.9 & 12.8$\pm$1.8 & 3.82$\pm$0.08 \\ 
B12-067 & 119.2$\pm$2.4 & 61.5$\pm$3.1 & 150.0$\pm$3.4 & 43.3$\pm$2.6 & 300 & 288.7$\pm$6.0 & 151.5$\pm$3.2 & 122.0$\pm$3.0 & 18.7$\pm$3.2 & 4.06$\pm$0.07 \\ 
B12-066 & 78.8$\pm$1.6 & 53.3$\pm$5.8 & 59.5$\pm$5.3 & 52.6$\pm$4.2 & 300 & 356.1$\pm$6.1 & 107.0$\pm$3.0 & 116.7$\pm$3.0 & 20.7$\pm$4.4 & 4.32$\pm$0.07 \\ 
L22-006 & 8.8$\pm$0.4 & 53.6$\pm$18.3 & 79.1$\pm$18.6 & -- & 300 & 247.2$\pm$13.7 & 170.1$\pm$10.9 & 127.3$\pm$10.3 & 0.0$\pm$13.1 & 2.88$\pm$0.13 \\ 
B12-069 & 27.2$\pm$0.6 & 73.2$\pm$9.6 & 183.1$\pm$10.6 & 41.0$\pm$8.3 & 300 & 268.9$\pm$6.2 & 130.6$\pm$5.5 & 91.5$\pm$5.2 & 15.0$\pm$6.8 & 3.18$\pm$0.06 \\ 
L22-007 & 24.2$\pm$0.5 & 51.2$\pm$6.6 & 93.3$\pm$6.1 & 34.6$\pm$4.6 & 300 & 238.2$\pm$5.8 & 119.5$\pm$3.3 & 74.4$\pm$3.0 & 7.7$\pm$2.9 & 3.61$\pm$0.07 \\ 
B12-070 & 27.1$\pm$0.5 & 83.6$\pm$16.8 & 112.1$\pm$11.7 & 44.8$\pm$8.3 & 300 & 260.2$\pm$5.3 & 142.7$\pm$5.2 & 107.4$\pm$5.0 & 15.1$\pm$9.1 & 3.25$\pm$0.05 \\ 
B12-071 & 76.0$\pm$1.1 & 62.3$\pm$5.6 & 99.3$\pm$5.6 & 64.3$\pm$4.5 & 300 & 366.6$\pm$4.7 & 180.4$\pm$3.4 & 127.8$\pm$3.2 & 17.1$\pm$4.6 & 4.92$\pm$0.06 \\ 
L22-008 & 196.8$\pm$1.3 & 60.8$\pm$1.8 & 7.8$\pm$1.1 & 12.6$\pm$1.1 & 300 & 121.7$\pm$1.8 & 55.4$\pm$0.9 & 41.6$\pm$0.9 & 8.7$\pm$1.1 & 2.81$\pm$0.02 \\ 
L22-009 & 8.2$\pm$0.2 & 53.4$\pm$15.9 & 119.1$\pm$14.5 & 82.9$\pm$11.8 & 300 & 338.1$\pm$9.0 & 160.3$\pm$7.4 & 115.3$\pm$6.9 & 0.0$\pm$8.0 & 4.33$\pm$0.11 \\ 
B12-074 & 32.8$\pm$0.5 & 63.9$\pm$8.0 & 301.6$\pm$5.0 & 97.2$\pm$4.3 & 300 & 589.5$\pm$5.2 & 174.7$\pm$3.9 & 139.1$\pm$3.7 & 0.0$\pm$2.7 & 5.01$\pm$0.04 \\ 
B12-075 & 250.5$\pm$6.5 & 61.7$\pm$2.1 & 87.1$\pm$1.5 & 66.0$\pm$1.1 & 300 & 266.2$\pm$7.6 & 131.0$\pm$2.1 & 127.2$\pm$2.1 & 19.9$\pm$1.2 & 4.35$\pm$0.10 \\ 
B12-076 & 20.8$\pm$0.5 & 59.6$\pm$13.8 & 36.8$\pm$8.5 & 74.6$\pm$7.3 & 300 & 251.0$\pm$6.7 & 133.7$\pm$6.2 & 119.7$\pm$6.0 & 10.9$\pm$7.3 & 3.94$\pm$0.08 \\ 
\enddata
\tablenotetext{a}{Surface brightness in units of 10$^{-17}$ ergs cm$^{-2}$ s$^{-1}$ arcsec$^{-2}$}
\tablenotetext{b}{Ratio to H$\alpha$  flux where, by convention, H$\alpha$ is normalized to 300.}
\tablenotetext{c}{In units of \AA.}
\tablecomments{This table with 229 rows is available in its entirety in machine-readable form}
\label{snr_spectra}
\end{deluxetable}

\pagebreak
\begin{deluxetable}{lrrrrc}
\tabletypesize{\scriptsize}
\decimals
\tablecaption{Comparison HII Regions in M83}
\tablehead{
\colhead{Source name} &
\colhead{RA} &
\colhead{Dec} &
\colhead{D} &
\colhead{R}
\\
\colhead{} &
\colhead{(J2000)} &
\colhead{(J2000)} &
\colhead{(pc)} &
\colhead{(kpc)}
}
\startdata
HII-001 & 204.20980 & -29.85976 & 30 & 3.3 \\ 
HII-002 & 204.21289 & -29.85073 & 30 & 3.4 \\ 
HII-003 & 204.21551 & -29.85977 & 19 & 2.9 \\ 
HII-004 & 204.21578 & -29.85089 & 26 & 3.1 \\ 
HII-005 & 204.21589 & -29.86071 & 31 & 2.9 \\ 
HII-006 & 204.21708 & -29.86958 & 27 & 2.7 \\ 
HII-007 & 204.21870 & -29.87038 & 23 & 2.6 \\ 
HII-008 & 204.21872 & -29.86566 & 46 & 2.6 \\ 
HII-009 & 204.21901 & -29.86967 & 28 & 2.5 \\ 
HII-010 & 204.22009 & -29.87208 & 36 & 2.5 \\ 
HII-011 & 204.22061 & -29.87013 & 18 & 2.4 \\ 
HII-012 & 204.22193 & -29.86755 & 24 & 2.3 \\ 
HII-013 & 204.22197 & -29.87031 & 40 & 2.3 \\ 
HII-014 & 204.22315 & -29.84693 & 39 & 2.9 \\ 
HII-015 & 204.22358 & -29.87392 & 30 & 2.3 \\ 
HII-016 & 204.22409 & -29.86566 & 27 & 2.2 \\ 
HII-017 & 204.22442 & -29.87589 & 22 & 2.3 \\ 
HII-018 & 204.22455 & -29.86160 & 50 & 2.2 \\ 
HII-019 & 204.22643 & -29.86068 & 37 & 2.1 \\ 
HII-020 & 204.22714 & -29.85770 & 34 & 2.1 \\ 
HII-021 & 204.22726 & -29.87350 & 29 & 2.0 \\ 
HII-022 & 204.22782 & -29.89359 & 37 & 2.9 \\ 
HII-023 & 204.22821 & -29.87708 & 29 & 2.0 \\ 
HII-024 & 204.22863 & -29.85969 & 86 & 2.0 \\ 
HII-025 & 204.22939 & -29.85573 & 33 & 2.0 \\ 
HII-026 & 204.22963 & -29.86719 & 26 & 1.8 \\ 
HII-027 & 204.23233 & -29.84877 & 25 & 2.2 \\ 
HII-028 & 204.23259 & -29.83931 & 23 & 2.8 \\ 
HII-029 & 204.23284 & -29.86544 & 22 & 1.5 \\ 
HII-030 & 204.23351 & -29.87683 & 23 & 1.7 \\ 
HII-031 & 204.23388 & -29.86347 & 26 & 1.5 \\ 
HII-032 & 204.23412 & -29.87030 & 36 & 1.5 \\ 
HII-033 & 204.23531 & -29.87662 & 35 & 1.6 \\ 
HII-034 & 204.23574 & -29.84953 & 28 & 2.0 \\ 
HII-035 & 204.23624 & -29.85967 & 33 & 1.4 \\ 
HII-036 & 204.23625 & -29.87032 & 36 & 1.3 \\ 
HII-037 & 204.23697 & -29.88843 & 32 & 2.2 \\ 
HII-038 & 204.23699 & -29.88010 & 67 & 1.7 \\ 
HII-039 & 204.23702 & -29.86251 & 26 & 1.3 \\ 
HII-040 & 204.23703 & -29.85281 & 27 & 1.7 \\ 
\enddata
\tablecomments{This table with 188 rows is available in its entirety in machine-readable form}
\label{h2_master}
\end{deluxetable}

\pagebreak
\begin{deluxetable}{rrrrrrrrrrr}
\tabletypesize{\scriptsize}
\decimals
\tablecaption{MUSE spectra of M83 HII regions}
\tablehead{
\colhead{Source} &
\colhead{H$\alpha$ SB$^a$} &
\colhead{H$\beta^b$} &
\colhead{[OIII]5007$^b$} &
\colhead{[OI]6300$^b$} &
\colhead{H$\alpha$} &
\colhead{[NII]6584$^b$} &
\colhead{[SII]6716$^b$} &
\colhead{[SII]6731$^b$} &
\colhead{[SIII]9069$^b$} &
\colhead{H$\alpha$ FWHM$^c$}
}
\startdata
HII-001 & 54.4$\pm$0.4 & 59.2$\pm$5.5 & 0.0$\pm$3.2 & 4.9$\pm$14.6 & 300 & 66.5$\pm$1.6 & 27.3$\pm$1.8 & 21.7$\pm$1.7 & 7.2$\pm$3.2 & 2.40$\pm$0.02 \\ 
HII-002 & 40.3$\pm$0.4 & 64.8$\pm$4.0 & 0.0$\pm$3.9 & 23.8$\pm$3.1 & 300 & 90.9$\pm$2.6 & 33.8$\pm$2.5 & 22.1$\pm$2.3 & 14.8$\pm$3.4 & 2.53$\pm$0.03 \\ 
HII-003 & 47.0$\pm$0.4 & 66.8$\pm$4.8 & 5.7$\pm$3.1 & 0.0$\pm$4.7 & 300 & 103.0$\pm$2.2 & 52.6$\pm$1.9 & 40.0$\pm$1.8 & 7.6$\pm$2.9 & 2.65$\pm$0.02 \\ 
HII-004 & 38.7$\pm$0.4 & 32.3$\pm$4.2 & 33.1$\pm$8.6 & 0.0$\pm$5.7 & 300 & 138.1$\pm$3.0 & 46.0$\pm$2.3 & 38.5$\pm$2.2 & 28.7$\pm$3.2 & 2.79$\pm$0.03 \\ 
HII-005 & 55.0$\pm$0.3 & 58.1$\pm$3.4 & 15.3$\pm$3.6 & 8.8$\pm$1.7 & 300 & 131.7$\pm$1.6 & 39.1$\pm$1.6 & 29.1$\pm$1.5 & 17.7$\pm$2.0 & 2.58$\pm$0.02 \\ 
HII-006 & 23.4$\pm$0.2 & 65.6$\pm$5.7 & 11.0$\pm$5.4 & 0.0$\pm$8.3 & 300 & 135.4$\pm$2.7 & 35.8$\pm$2.4 & 25.0$\pm$2.2 & 0.0$\pm$4.0 & 2.74$\pm$0.03 \\ 
HII-007 & 92.0$\pm$0.4 & 67.2$\pm$1.7 & 4.5$\pm$1.1 & 3.2$\pm$1.3 & 300 & 109.3$\pm$1.0 & 23.4$\pm$0.8 & 17.3$\pm$0.7 & 8.6$\pm$1.2 & 2.63$\pm$0.01 \\ 
HII-008 & 94.8$\pm$0.3 & 45.1$\pm$1.3 & 4.8$\pm$1.2 & 1.2$\pm$0.9 & 300 & 100.3$\pm$0.9 & 32.1$\pm$0.7 & 23.5$\pm$0.6 & 11.8$\pm$0.8 & 2.62$\pm$0.01 \\ 
HII-009 & 31.3$\pm$0.2 & 78.8$\pm$5.4 & 14.7$\pm$4.8 & 11.1$\pm$2.6 & 300 & 178.7$\pm$2.1 & 49.3$\pm$2.1 & 34.0$\pm$2.0 & 0.0$\pm$3.0 & 2.61$\pm$0.02 \\ 
HII-010 & 251.5$\pm$0.9 & 50.0$\pm$0.4 & 8.0$\pm$0.4 & 1.5$\pm$0.3 & 300 & 149.2$\pm$1.0 & 30.8$\pm$0.3 & 22.2$\pm$0.3 & 20.8$\pm$0.4 & 2.56$\pm$0.01 \\ 
HII-011 & 9.4$\pm$0.3 & 0.0$\pm$14.5 & 0.0$\pm$12.4 & 43.6$\pm$11.0 & 300 & 207.6$\pm$9.3 & 95.3$\pm$7.3 & 56.6$\pm$6.6 & 0.0$\pm$11.4 & 2.95$\pm$0.10 \\ 
HII-012 & 23.9$\pm$0.2 & 63.7$\pm$6.5 & 0.0$\pm$4.0 & 9.1$\pm$8.5 & 300 & 122.0$\pm$2.5 & 40.3$\pm$3.0 & 25.0$\pm$2.8 & 0.0$\pm$4.8 & 2.60$\pm$0.03 \\ 
HII-013 & 363.5$\pm$1.1 & 71.0$\pm$0.5 & 8.6$\pm$0.3 & 1.2$\pm$0.2 & 300 & 122.1$\pm$0.8 & 22.6$\pm$0.3 & 16.2$\pm$0.3 & 15.6$\pm$0.3 & 2.51$\pm$0.01 \\ 
HII-014 & 46.1$\pm$0.4 & 59.5$\pm$5.3 & 67.4$\pm$4.3 & 18.2$\pm$2.6 & 300 & 115.5$\pm$2.4 & 45.9$\pm$2.3 & 26.0$\pm$2.0 & 27.7$\pm$2.3 & 2.53$\pm$0.03 \\ 
HII-015 & 22.7$\pm$0.4 & 63.4$\pm$8.9 & 10.2$\pm$5.7 & 57.1$\pm$6.1 & 300 & 79.5$\pm$4.0 & 37.0$\pm$3.7 & 25.1$\pm$3.4 & 0.0$\pm$5.3 & 2.75$\pm$0.05 \\ 
HII-016 & 65.0$\pm$0.4 & 54.1$\pm$2.4 & 6.7$\pm$2.0 & 0.0$\pm$3.0 & 300 & 89.3$\pm$1.5 & 28.7$\pm$1.2 & 20.7$\pm$1.1 & 11.4$\pm$1.5 & 2.51$\pm$0.02 \\ 
HII-017 & 33.3$\pm$0.5 & 37.3$\pm$6.8 & 0.0$\pm$5.3 & 17.0$\pm$7.0 & 300 & 98.8$\pm$3.6 & 33.9$\pm$3.2 & 25.0$\pm$3.1 & 0.0$\pm$5.7 & 2.56$\pm$0.04 \\ 
HII-018 & 7.6$\pm$0.2 & 0.0$\pm$11.8 & 190.1$\pm$16.1 & 12.1$\pm$14.2 & 300 & 212.2$\pm$7.6 & 82.9$\pm$7.6 & 52.3$\pm$6.9 & 70.7$\pm$14.2 & 2.47$\pm$0.06 \\ 
HII-019 & 59.6$\pm$0.3 & 68.2$\pm$2.8 & 8.0$\pm$1.8 & 4.7$\pm$1.3 & 300 & 159.3$\pm$1.3 & 40.3$\pm$0.9 & 25.1$\pm$0.8 & 18.7$\pm$1.7 & 2.66$\pm$0.01 \\ 
HII-020 & 8.7$\pm$0.2 & 75.1$\pm$11.6 & 0.0$\pm$9.4 & 22.4$\pm$4.8 & 300 & 103.0$\pm$5.6 & 27.1$\pm$5.8 & 17.6$\pm$5.3 & 0.0$\pm$7.7 & 2.58$\pm$0.06 \\ 
HII-021 & 55.0$\pm$0.4 & 57.8$\pm$2.3 & 7.0$\pm$2.4 & 4.5$\pm$1.5 & 300 & 137.9$\pm$1.7 & 35.8$\pm$1.4 & 25.1$\pm$1.3 & 8.8$\pm$2.0 & 2.62$\pm$0.02 \\ 
HII-022 & 200.9$\pm$0.9 & 73.8$\pm$1.4 & 4.3$\pm$1.0 & 1.6$\pm$0.7 & 300 & 76.9$\pm$1.1 & 28.7$\pm$0.6 & 19.2$\pm$0.5 & 7.5$\pm$0.9 & 2.55$\pm$0.01 \\ 
HII-023 & 12.4$\pm$0.3 & 49.0$\pm$13.4 & 38.7$\pm$19.4 & 0.0$\pm$17.4 & 300 & 129.4$\pm$6.7 & 62.2$\pm$7.2 & 45.1$\pm$6.8 & 0.0$\pm$8.6 & 2.74$\pm$0.08 \\ 
HII-024 & 4.7$\pm$0.2 & 0.0$\pm$15.4 & 0.0$\pm$14.0 & 0.0$\pm$40.3 & 300 & 83.5$\pm$9.7 & 62.4$\pm$8.7 & 37.1$\pm$7.9 & 0.0$\pm$13.2 & 2.61$\pm$0.12 \\ 
HII-025 & 5.8$\pm$0.2 & 30.7$\pm$13.2 & 0.0$\pm$12.9 & 29.6$\pm$13.4 & 300 & 82.0$\pm$8.1 & 62.0$\pm$8.5 & 46.1$\pm$8.0 & 0.0$\pm$11.3 & 2.49$\pm$0.09 \\ 
HII-026 & 10.4$\pm$0.3 & 142.8$\pm$25.2 & 45.9$\pm$12.5 & 30.7$\pm$14.6 & 300 & 175.1$\pm$6.9 & 64.2$\pm$7.5 & 53.0$\pm$7.2 & 0.0$\pm$10.1 & 2.60$\pm$0.07 \\ 
HII-027 & 79.3$\pm$0.5 & 54.8$\pm$2.5 & 70.7$\pm$1.9 & 17.2$\pm$1.5 & 300 & 106.6$\pm$1.5 & 28.1$\pm$1.3 & 22.6$\pm$1.2 & 28.6$\pm$1.8 & 2.61$\pm$0.02 \\ 
HII-028 & 48.4$\pm$0.4 & 65.4$\pm$4.7 & 15.0$\pm$4.9 & 0.0$\pm$6.6 & 300 & 137.2$\pm$2.4 & 43.4$\pm$2.0 & 34.6$\pm$1.9 & 22.9$\pm$2.5 & 2.68$\pm$0.03 \\ 
HII-029 & 13.1$\pm$0.3 & 84.0$\pm$11.3 & 16.2$\pm$7.7 & 21.0$\pm$8.1 & 300 & 107.2$\pm$5.1 & 43.0$\pm$5.3 & 38.3$\pm$5.2 & 0.0$\pm$7.6 & 2.49$\pm$0.06 \\ 
HII-030 & 53.5$\pm$0.6 & 92.6$\pm$6.4 & 90.7$\pm$4.4 & 12.0$\pm$4.3 & 300 & 210.1$\pm$3.0 & 37.5$\pm$2.8 & 31.9$\pm$2.7 & 30.6$\pm$3.9 & 2.65$\pm$0.03 \\ 
HII-031 & 7.5$\pm$0.2 & 92.7$\pm$13.5 & 628.3$\pm$14.0 & 36.0$\pm$14.0 & 300 & 107.4$\pm$8.0 & 31.3$\pm$6.6 & 50.1$\pm$7.2 & 0.0$\pm$10.1 & 3.13$\pm$0.11 \\ 
HII-032 & 66.1$\pm$0.3 & 68.5$\pm$2.5 & 3.1$\pm$2.0 & 5.0$\pm$1.5 & 300 & 99.9$\pm$1.3 & 29.1$\pm$1.0 & 19.6$\pm$1.0 & 7.0$\pm$1.4 & 2.61$\pm$0.01 \\ 
HII-033 & 5.7$\pm$0.6 & 450.3$\pm$133.5 & 0.0$\pm$33.1 & 0.0$\pm$40.0 & 300 & 310.4$\pm$31.8 & 97.9$\pm$29.7 & 37.0$\pm$25.6 & 0.0$\pm$34.4 & 3.30$\pm$0.30 \\ 
HII-034 & 24.6$\pm$0.2 & 73.4$\pm$5.4 & 0.0$\pm$3.8 & 0.0$\pm$8.3 & 300 & 103.7$\pm$2.4 & 49.8$\pm$2.0 & 29.4$\pm$1.8 & 9.7$\pm$3.7 & 2.70$\pm$0.03 \\ 
HII-035 & 28.7$\pm$0.2 & 69.5$\pm$3.2 & 0.0$\pm$2.1 & 0.0$\pm$6.3 & 300 & 62.5$\pm$1.8 & 19.8$\pm$1.5 & 10.3$\pm$1.4 & 8.6$\pm$1.6 & 2.49$\pm$0.02 \\ 
HII-036 & 35.6$\pm$1.3 & 80.4$\pm$4.3 & 30.6$\pm$6.9 & 9.4$\pm$4.0 & 300 & 324.4$\pm$11.3 & 28.6$\pm$1.9 & 21.3$\pm$1.8 & 0.0$\pm$2.5 & 5.40$\pm$0.17 \\ 
HII-037 & 70.1$\pm$0.5 & 80.8$\pm$3.4 & 0.0$\pm$2.4 & 0.0$\pm$3.1 & 300 & 65.1$\pm$1.8 & 22.5$\pm$1.5 & 15.1$\pm$1.3 & 0.0$\pm$2.2 & 2.62$\pm$0.02 \\ 
HII-038 & 1571.5$\pm$4.5 & 37.8$\pm$0.2 & 4.6$\pm$0.2 & 0.7$\pm$0.1 & 300 & 124.1$\pm$0.7 & 22.7$\pm$0.3 & 17.5$\pm$0.3 & 40.5$\pm$0.1 & 2.52$\pm$0.01 \\ 
HII-039 & 132.7$\pm$0.3 & 66.3$\pm$0.8 & 3.2$\pm$0.5 & 2.3$\pm$0.5 & 300 & 116.8$\pm$0.6 & 23.7$\pm$0.3 & 16.3$\pm$0.3 & 14.7$\pm$0.4 & 2.49$\pm$0.01 \\ 
HII-040 & 7.5$\pm$0.2 & 65.2$\pm$13.0 & 0.0$\pm$9.3 & 0.0$\pm$25.4 & 300 & 144.1$\pm$7.3 & 51.0$\pm$6.2 & 31.4$\pm$5.9 & 0.0$\pm$9.5 & 2.53$\pm$0.08 \\ 
\enddata
\tablenotetext{a}{Surface brightness in units of 10$^{-17}$ ergs cm$^{-2}$ s$^{-1}$ arcsec$^{-2}$}
\tablenotetext{b}{Ratio to H$\alpha$  flux where, by convention, H$\alpha$ is normalized to 300.}
\tablenotetext{c}{In units of \AA.}
\tablecomments{This table with 188 rows is available in its entirety in machine-readable form}
\label{h2_spectra}
\end{deluxetable}

\end{document}